\documentclass[twocolumn,twocolappendix]{aastex631}
\usepackage{graphicx}	% Including figure files
\usepackage{amsmath}	% Advanced maths commands

\begin{document}

\title[The particularity of the active episode of SGR J1935+2154 during which FRB 20200428 occurred]{Finding the Particularity of the Active Episode of SGR J1935+2154 during Which FRB 20200428 Occurred: Implication from Statistics of Fermi/GBM X-Ray Bursts}

% Author
\author[0000-0001-9217-7070]{Sheng-Lun Xie}
\affiliation{Institute of Astrophysics, Central China Normal University, Wuhan 430079, China}
\affiliation{Key Laboratory of Particle Astrophysics, Institute of High Energy Physics, Chinese Academy of Sciences, 19B Yuquan Road, Beijing 100049, China}
% \email{xiesl@mails.ccnu.edu.cn}
%
\author[0000-0002-1067-1911]{Yun-Wei Yu}
\affiliation{Institute of Astrophysics, Central China Normal University, Wuhan 430079, China}
\affiliation{Key Laboratory of Quark and Lepton Physics (Central China Normal University), 
Ministry of Education, Wuhan 430079, China}
\correspondingauthor{Yun-Wei Yu}
\email{yuyw@ccnu.edu.cn}
\author{Shao-Lin Xiong}
\affiliation{Key Laboratory of Particle Astrophysics, Institute of High Energy Physics, Chinese Academy of Sciences, 19B Yuquan Road, Beijing 100049, China}
\correspondingauthor{Shao-Lin Xiong}
\email{xiongsl@ihep.ac.cn}
\author{Lin Lin}
\affiliation{Department of Astronomy, Beijing Normal University, Beijing 100088, China}
% \email{llin@bnu.edu.cn}
%
\author{Ping Wang}
\affiliation{Key Laboratory of Particle Astrophysics, Institute of High Energy Physics, Chinese Academy of Sciences, 19B Yuquan Road, Beijing 100049, China}
% \email{pwang@ihep.ac.cn}
%
\author{Yi Zhao}
\affiliation{School of Computer and Information, Dezhou University, Dezhou 253023, China}
% \email{yizhao@dzu.edu.cn}
%
\author{Yue Wang}
\affiliation{Key Laboratory of Particle Astrophysics, Institute of High Energy Physics, Chinese Academy of Sciences, 19B Yuquan Road, Beijing 100049, China}
% \email{yuewang@ihep.ac.cn}
%
\author{Wen-Long Zhang}
\affiliation{School of Physics and Physical Engineering, Qufu Normal University, Qufu, Shandong 273165, China}
\affiliation{Key Laboratory of Particle Astrophysics, Institute of High Energy Physics, Chinese Academy of Sciences, 19B Yuquan Road, Beijing 100049, China}
% \email{zhangwl@ihep.ac.cn}

% Abstract
\begin{abstract}
By using the Fermi/Gamma-ray Burst Monitor data of the X-ray bursts (XRBs) of SGR J1935+2154, we investigate the temporal clustering of the bursts and the cumulative distribution of the waiting time and fluence/flux. It is found that the bursts occurring in the episode hosting FRB 20200428 have obviously shorter waiting times than those in the other episodes. The general statistical properties of the XRBs further indicate they could belong to a self-organized critical (SOC) system (e.g., starquakes), making them very similar to the earthquake phenomena. Then, according to a unified scaling law between the waiting time and energy of the earthquakes as well as their aftershocks, we implement an analogy analysis on the XRBs and find that the FRB episode owns more dependent burst events than the other episodes. It is indicated that the fast radio burst (FRB) emission could be produced by the interaction between different burst events, which could correspond to a collision between different seismic/Alfven waves or different explosion outflows. Such a situation could appear when the magnetar enters into a global intensive activity period.
\end{abstract}


\section{Introduction} \label{sec:intro}

Generally, magnetars are a class of neutron stars with an ultrahigh magnetic field ($\sim 10^{14}-10^{15}$ G), which usually exhibit in the Galaxy as pulsars of a long spin period ($2\sim 12$ s) and a fast spin-down rate \citep[$\sim 10^{-13}-10^{-11}$ s s$^{-1}$, ][]{Kouveliotou1998,Olausen2014}. The marked feature of Galactic magnetars is their X-ray or soft $\gamma$-ray burst emission \citep{ThompsonDuncan1995}, which are observed in more than 2/3 of the magnetar population \citep{Olausen2014}.

On 28th April 2020, from the direction of a Galactic magnetar, SGR J1935+2154, the Canadian Hydrogen Intensity Mapping Experiment (CHIME) captured a fast radio burst (FRB) emission \citep{Bochenek2020,CHIME2020b}, which was precisely temporally associated with an X-ray burst \citep[XRB;][]{Li2021,LiXiaobo2022,Ge2022arxiv}. This magnetar had just entered an active state of XRBs about half a month before FRB 20200428 occurred \citep{Li2021,Cai2022a,Cai2022b,Mereghetti2020,Younes2020,Tavani2020,Veres2020,Ridnaia2020,Barthelmy2020,Lin2020b,Ridnaia2021,Kaneko2021}, and remained active at other times as well \citep{Israel2016MNRAS,Lin2020a,Borghese2022MNRAS,Xie2022,Xie2023arxiv,Rehan2023}. Meanwhile, the dispersion measure (DM) of FRB 20200428 ($\sim332.7$ pc cm $^{-3}$) is well consistent with the distance of SGR J1935+2154 in the range of $6.6-12.5$ kpc \citep{Kothes2018,Zhou2020,Zhong2020}. These discoveries indicated that magnetars could be sources of the mysterious FRB phenomena \citep{Lorimer2007,Thornton2013}.

In view of their millisecond duration and high energy releases, FRBs are widely suggested to originate from the violent activities of compact objects, in particular, magnetars \citep{Popov2010,Kulkarni2014,Katz2016,Connor2016,Cordes2016,Lyutikov2017}, which was firstly favored by the discovery of the repeating FRB 20121102 as well as its persistent radio counterpart \citep{Kashiyama2017,Metzger2017,Cao2017,Dai2017,Michilli2018}. Then, the direct connection of FRB 20200428 with SGR J1935+2154 undoubtedly offered a smoking-gun evidence for one of the FRB/FRB-like origins, the magnetar origin, although FRB 20200428 is still distinctive due to its obviously low luminosity than its cosmological cousins \citep{CHIME2020b,Bochenek2020}.

Generally, the amounts of the energy releases of the bursts and outbursts of Galactic magnetars, from both soft gamma-ray repeaters (SGRs) and anomalous X-ray pulsars (AXPs), indicate that XRBs probably correspond to sudden and local (not global) catastrophic events. Since the energy is considered to eventually come from magnetic fields, the XRBs are most likely to be caused by sudden reordering of the magnetic field structure, which either deforms the crust drastically or/and leads to a major reconfiguration of the magnetosphere. As a result, an optically thick $e^{\pm}$ fireball can be expected to form, radiating X-rays \citep{ThompsonDuncan1995,ThompsonDuncan2001,Beloborodov2021ApJ,Lyutikov2022MNRAS}. In the framework of such a starquake model, regardless of the specific X-ray radiation mechanism, some scale independent behaviors can be expected to appear in the statistics of the energy and waiting times of XRBs, just as found from earthquakes \citep{Cheng1996}. Many systems such as starquakes and earthquakes, consisting of a large number of entities that interact with each other in a complex way, can exhibit nonlinear behavior, which can be described by a self-organized criticality (SOC) theory. The complex interactions in such systems can occasionally lead to instabilities with subsequent small events. The largest events, the subsequent events, and the myriads of smaller events can finally share the same statistical property \citep{Sornette2004book,Tsallis2009book,Aschwanden2011book}.

Despite a basic consensus that could appear on the XRB origin, the mechanism responsible for the FRB emission is still under great debate. In the literature, two types of models have been suggested frequently, including the curvature radiation in the magnetosphere \citep{LuKumar2018MNRAS,YangZhang2018ApJ,YangZhang2020ApJ,LongPeer2018ApJ,WangZhang2019ApJ,KumarLu2020MNRAS} and the shock synchrotron maser radiation \citep{Lyubarsky2014MNRAS,Beloborodov2017ApJ,Ghisellini2017MNRAS,KumarLu2017MNRAS,Lyutikov2017ApJ,Waxman2017ApJ,Metzger2019MNRAS,Plotnikov2019MNRAS,YuZou2021MNRAS}. In the former case, the curvature radiation could happen in the magnetosphere because the crust fracturing forms a charge-starved region and leads to the plasma discharges. In the latter case, the shock responsible for the synchrotron maser can be driven by the collisions between different ejecta of different burst activities. In observation, although FRB 20200428 was found to be in temporal coincidence with an XRB, the deep observations implemented by the Five Hundred Meter Aperture Telescope (FAST) on some other XRBs of SGR J1935+2154 ruled out the possibility of their association with FRB emission \citep{Lin2020c}. On the one hand, this fact could just be due to the narrow beaming of FRB emission, while the XRB emission is generally isotropic. In this case, there is in principle no intrinsic difference between the different XRBs. On the other hand, reversely, the un-association with FRB emission of other XRBs may indicate that the XRB associated with FRB 20200428 is intrinsically different from the others.

Therefore, in this paper, we first investigate the statistical distributions of the energy and waiting times of the XRBs of SGR J1935+2154 in order to reveal their possible SOC behaviors. Furthermore, these XRBs are separated into several different active episodes and the possible particularity of the active episode where FRB 20200428 occurred is further investigated, in order to test whether the FRB-associated XRB is intrinsically different from the other XRBs or the non-detection of FRB emission is just because of the off-axis observational direction. This could help us to uncover the FRB emission mystery.

\begin{figure}
\centering
\centerline{\includegraphics[width=0.45\textwidth]{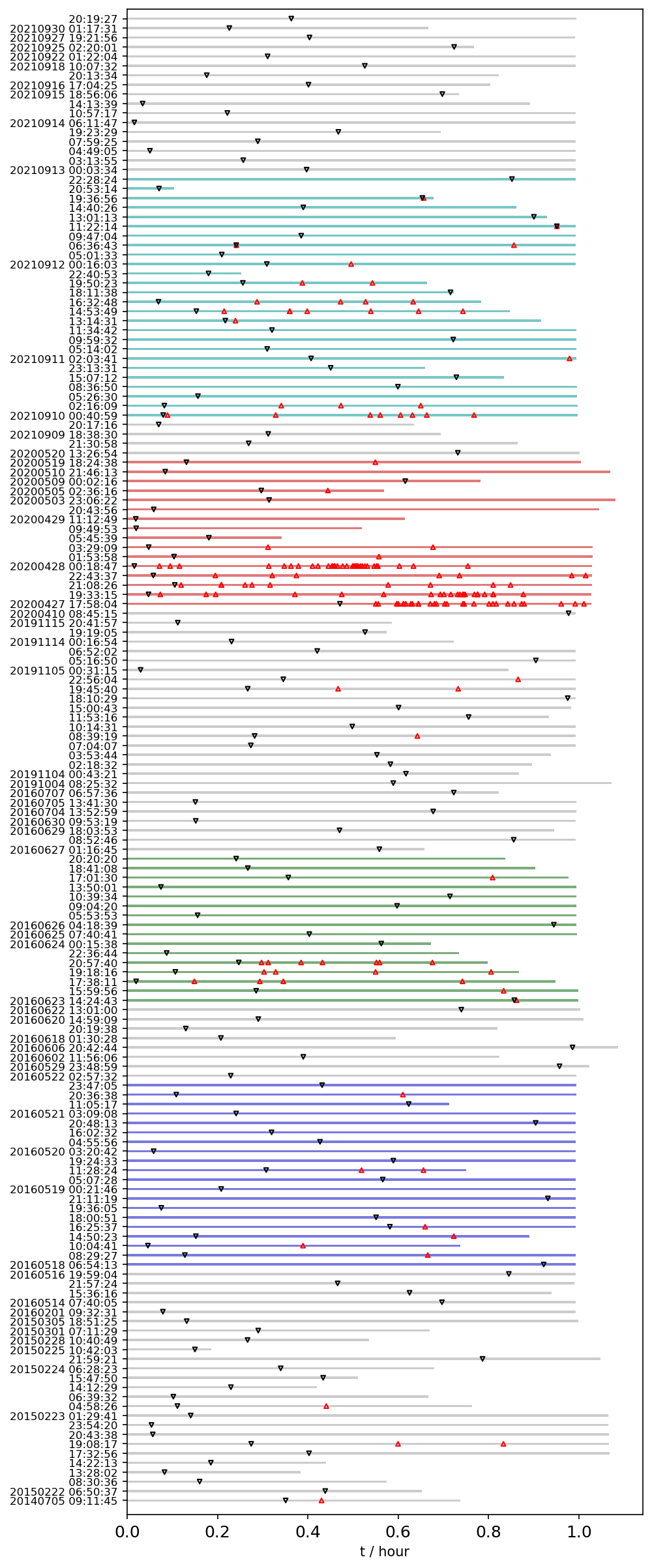}}
\caption{
The XRBs (triangle markers) of SGR J1935+2154 detected by GBM within different observation windows (horizontal lines). The length of the lines represents the duration of the windows. The first burst in each window is labelled by a black triangle, which is dropped in our statistics since its waiting time cannot be defined. Then, the windows that only have one burst are excluded (i.e, the grey lines). The remaining windows having sufficiently large number of bursts can be basically separated into four active episodes, as labelled by different colors (see Section \ref{subsec:cd} for active episode definition).
}\label{fig:ObsWindow}
\end{figure}

\begin{figure}
\centering
\centerline{\includegraphics[width=0.45\textwidth]{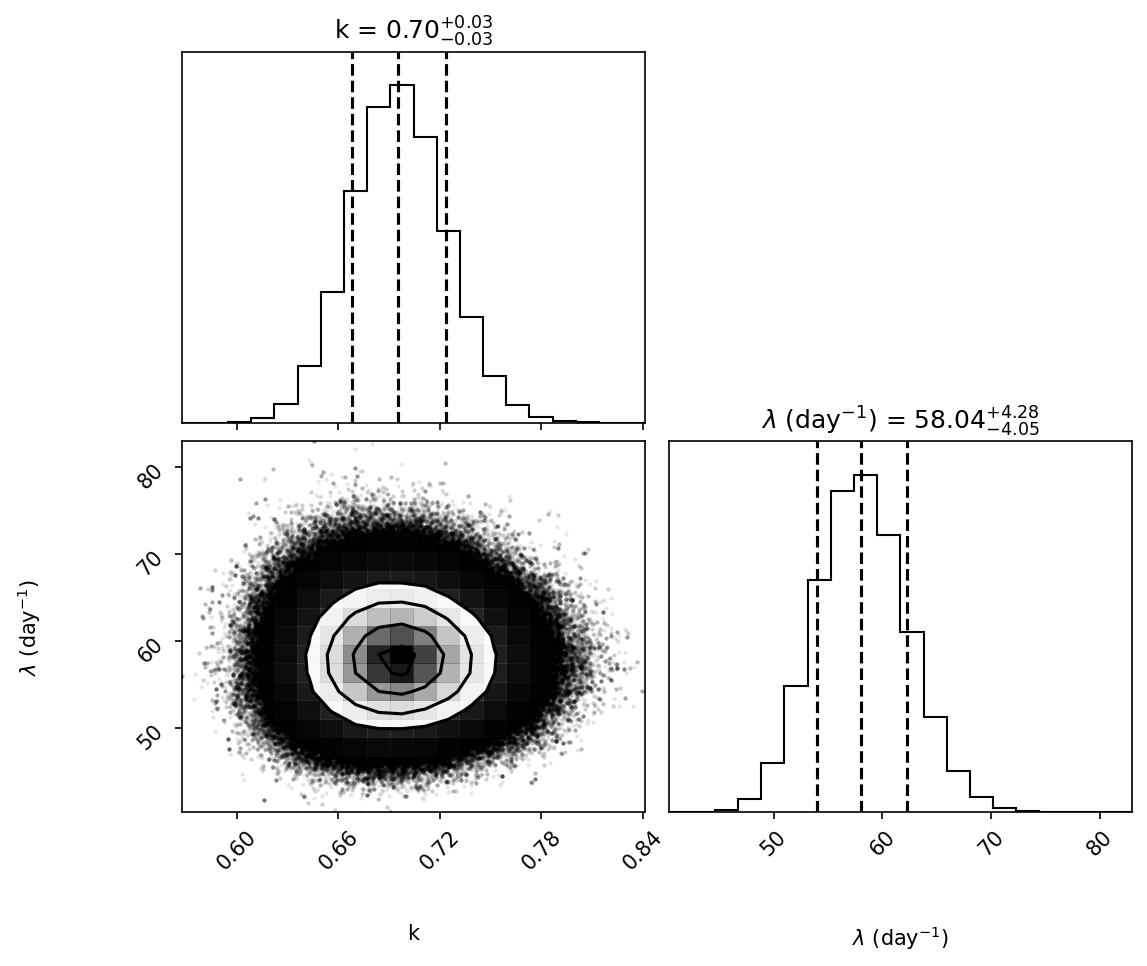}}
\caption{The corner plot of the shape parameter $k$ and the event rate $\lambda$ for the Weibull distribution
(see Section \ref{subsec:tc}).}\label{fig:mcmc}
\end{figure}

\begin{table*}
\scriptsize
\centering
\caption{The start time and duration $T$ of the observation windows of the \textit{Fermi}/GBM on SGR J1935+2154}
\label{tab:obs_win}
\begin{tabular}{lc|lc|lc|lc}
\hline
Start time & $T$ & Start time & $T$ & Start time & $T$ & Start time & $T$ \\
(UTC) & (hour) & (UTC) & (hour) & (UTC) & (hour) & (UTC) & (hour) \\
\hline
2014-07-05T09:11:45.34 & 0.74 & 2016-05-20T20:48:13.14 & 0.99 & 2019-11-04T07:04:07.94 & 0.99 & 2021-09-10T23:13:31.74 & 0.66 \\
2015-02-22T06:50:37.34 & 0.65 & 2016-05-21T03:09:08.14 & 0.99 & 2019-11-04T08:39:19.94 & 0.99 & 2021-09-11T02:03:41.74 & 1.00 \\
2015-02-22T08:30:36.34 & 0.57 & 2016-05-21T11:05:17.14 & 0.71 & 2019-11-04T10:14:31.94 & 0.99 & 2021-09-11T05:14:02.74 & 1.00 \\
2015-02-22T13:28:02.34 & 0.38 & 2016-05-21T20:36:38.14 & 0.99 & 2019-11-04T11:53:16.94 & 0.93 & 2021-09-11T09:59:32.74 & 0.99 \\
2015-02-22T14:22:13.34 & 0.44 & 2016-05-21T23:47:05.14 & 0.99 & 2019-11-04T15:00:43.94 & 0.98 & 2021-09-11T11:34:42.74 & 0.99 \\
2015-02-22T17:32:56.34 & 1.07 & 2016-05-22T02:57:32.14 & 0.99 & 2019-11-04T18:10:29.94 & 0.99 & 2021-09-11T13:14:31.74 & 0.92 \\
2015-02-22T19:08:17.34 & 1.07 & 2016-05-29T23:48:59.34 & 1.02 & 2019-11-04T19:45:40.94 & 0.99 & 2021-09-11T14:53:49.74 & 0.85 \\
2015-02-22T20:43:38.34 & 1.07 & 2016-06-02T11:56:06.74 & 0.82 & 2019-11-04T22:56:04.94 & 0.99 & 2021-09-11T16:32:48.74 & 0.78 \\
2015-02-22T23:54:20.34 & 1.07 & 2016-06-06T20:42:44.14 & 1.09 & 2019-11-05T00:31:15.94 & 0.84 & 2021-09-11T18:11:38.74 & 0.72 \\
2015-02-23T01:29:41.34 & 1.07 & 2016-06-18T01:30:28.54 & 0.59 & 2019-11-05T05:16:50.94 & 0.99 & 2021-09-11T19:50:23.74 & 0.66 \\
2015-02-23T04:58:26.34 & 0.76 & 2016-06-18T20:19:38.54 & 0.82 & 2019-11-05T06:52:02.94 & 0.99 & 2021-09-11T22:40:53.74 & 0.25 \\
2015-02-23T06:39:32.34 & 0.67 & 2016-06-20T14:59:09.54 & 1.01 & 2019-11-14T00:16:54.94 & 0.72 & 2021-09-12T00:16:03.74 & 0.99 \\
2015-02-23T14:12:29.34 & 0.42 & 2016-06-22T13:01:00.14 & 1.00 & 2019-11-14T19:19:05.94 & 0.58 & 2021-09-12T05:01:33.74 & 0.99 \\
2015-02-23T15:47:50.34 & 0.51 & 2016-06-23T14:24:43.14 & 1.00 & 2019-11-15T20:41:57.94 & 0.59 & 2021-09-12T06:36:43.74 & 0.99 \\
2015-02-24T06:28:23.34 & 0.68 & 2016-06-23T15:59:56.14 & 1.00 & 2020-04-10T08:45:15.54 & 0.99 & 2021-09-12T09:47:04.74 & 0.99 \\
2015-02-24T21:59:21.34 & 1.05 & 2016-06-23T17:38:11.14 & 0.95 & 2020-04-27T17:58:04.14 & 1.03 & 2021-09-12T11:22:14.74 & 0.99 \\
2015-02-25T10:42:03.34 & 0.19 & 2016-06-23T19:18:16.14 & 0.87 & 2020-04-27T19:33:15.14 & 1.03 & 2021-09-12T13:01:13.74 & 0.93 \\
2015-02-28T10:40:49.34 & 0.54 & 2016-06-23T20:57:40.14 & 0.80 & 2020-04-27T21:08:26.14 & 1.03 & 2021-09-12T14:40:26.74 & 0.86 \\
2015-03-01T07:11:29.34 & 0.67 & 2016-06-23T22:36:44.14 & 0.73 & 2020-04-27T22:43:37.14 & 1.03 & 2021-09-12T19:36:56.74 & 0.68 \\
2015-03-05T18:51:25.34 & 1.00 & 2016-06-24T00:15:38.14 & 0.67 & 2020-04-28T00:18:47.14 & 1.03 & 2021-09-12T20:53:14.74 & 0.10 \\
2016-02-01T09:32:31.14 & 0.99 & 2016-06-25T07:40:41.14 & 1.00 & 2020-04-28T01:53:58.14 & 1.03 & 2021-09-12T22:28:24.74 & 0.99 \\
2016-05-14T07:40:05.54 & 0.99 & 2016-06-26T04:18:39.54 & 1.00 & 2020-04-28T03:29:09.14 & 1.03 & 2021-09-13T00:03:34.74 & 0.99 \\
2016-05-14T15:36:16.54 & 0.94 & 2016-06-26T05:53:53.54 & 0.99 & 2020-04-28T05:45:39.14 & 0.34 & 2021-09-13T03:13:55.74 & 0.99 \\
2016-05-14T21:57:24.54 & 0.99 & 2016-06-26T09:04:20.54 & 0.99 & 2020-04-28T09:49:53.14 & 0.52 & 2021-09-13T04:49:05.74 & 0.99 \\
2016-05-16T19:59:04.54 & 0.99 & 2016-06-26T10:39:34.54 & 0.99 & 2020-04-29T11:12:49.14 & 0.61 & 2021-09-13T07:59:25.74 & 0.99 \\
2016-05-18T06:54:13.54 & 0.99 & 2016-06-26T13:50:01.54 & 0.99 & 2020-04-29T20:43:56.14 & 1.05 & 2021-09-13T19:23:29.74 & 0.69 \\
2016-05-18T08:29:27.54 & 0.99 & 2016-06-26T17:01:30.54 & 0.98 & 2020-05-03T23:06:22.14 & 1.08 & 2021-09-14T06:11:47.74 & 0.99 \\
2016-05-18T10:04:41.54 & 0.74 & 2016-06-26T18:41:08.54 & 0.90 & 2020-05-05T02:36:16.14 & 0.57 & 2021-09-14T10:57:17.74 & 0.99 \\
2016-05-18T14:50:23.54 & 0.89 & 2016-06-26T20:20:20.54 & 0.84 & 2020-05-09T00:02:16.14 & 0.78 & 2021-09-14T14:13:39.74 & 0.89 \\
2016-05-18T16:25:37.54 & 0.99 & 2016-06-27T01:16:45.54 & 0.66 & 2020-05-10T21:46:13.14 & 1.07 & 2021-09-15T18:56:06.74 & 0.74 \\
2016-05-18T18:00:51.54 & 0.99 & 2016-06-27T08:52:46.54 & 0.99 & 2020-05-19T18:24:38.14 & 1.00 & 2021-09-16T17:04:25.74 & 0.80 \\
2016-05-18T19:36:05.54 & 0.99 & 2016-06-29T18:03:53.54 & 0.95 & 2020-05-20T13:26:54.14 & 1.00 & 2021-09-16T20:13:34.74 & 0.82 \\
2016-05-18T21:11:19.54 & 0.99 & 2016-06-30T09:53:19.54 & 0.99 & 2020-05-20T21:30:58.94 & 0.86 & 2021-09-18T10:07:32.74 & 0.99 \\
2016-05-19T00:21:46.54 & 0.99 & 2016-07-04T13:52:59.54 & 0.99 & 2021-09-09T18:38:30.74 & 0.69 & 2021-09-22T01:22:04.74 & 0.99 \\
2016-05-19T05:07:28.54 & 0.99 & 2016-07-05T13:41:30.54 & 0.99 & 2021-09-09T20:17:16.74 & 0.63 & 2021-09-25T02:20:01.74 & 0.77 \\
2016-05-19T11:28:24.54 & 0.75 & 2016-07-07T06:57:36.54 & 0.82 & 2021-09-10T00:40:59.74 & 1.00 & 2021-09-27T19:21:56.74 & 0.99 \\
2016-05-19T19:24:33.14 & 0.99 & 2019-10-04T08:25:32.34 & 1.07 & 2021-09-10T02:16:09.74 & 1.00 & 2021-09-30T01:17:31.74 & 0.67 \\
2016-05-20T03:20:42.14 & 0.99 & 2019-11-04T00:43:21.94 & 0.87 & 2021-09-10T05:26:30.74 & 1.00 & 2021-09-30T20:19:27.74 & 1.00 \\
2016-05-20T04:55:56.14 & 0.99 & 2019-11-04T02:18:32.94 & 0.90 & 2021-09-10T08:36:50.74 & 1.00 &                        &      \\
2016-05-20T16:02:32.14 & 0.99 & 2019-11-04T03:53:44.94 & 0.94 & 2021-09-10T15:07:12.74 & 0.83 &                        &      \\
\hline
\end{tabular}
\end{table*}

%%%%%%%%%%%%%%%%%%%%%%%%%%%%%%%%%%%%%%%%%%%%%
\section{Temporal and energy distributions} \label{sec:tp}

We retrieve the burst start time, flux and fluence of SGR J1935+2154 which were detected by the \textit{Fermi}/Gamma-ray Burst Monitor (GBM) from the published burst history (2014-2021, Table 4 in \cite{Lin2020a}, Table 1 in \cite{Lin2020b}, Table 1 in \cite{Kaneko2021}, Table 1 in \cite{Rehan2023}). Then, we analyze the position history of Fermi/GBM observations using POSHIST \footnote{\url{https://heasarc.gsfc.nasa.gov/FTP/fermi/data/gbm/daily/}} and calculate the uninterrupted visible time (referred to as observation window hereafter) of the GBM to the SGR J1935+2154, excluding the Earth occultation and SAA (South Atlantic Anomaly) area, using the GBM Data Tools \citep{GbmDataTools}. The start times of the observation windows and hundreds of bursts are listed in Table \ref{tab:obs_win} and shown in Fig. \ref{fig:ObsWindow}. To be specific, the burst samples of SGR J1935+2154 consist of 158 observation windows, with each window having an approximate duration of an hour because GBM orbits the Earth every about 90 minutes. However some observation windows are noticeably shorter than the 90 minutes because GBM occasionally shuts down when the satellite crosses the SAA area.

\subsection{The temporal clustering of XRBs} \label{subsec:tc}

The waiting time between the successive bursts in the same observation window is defined as 
\begin{equation}
\Delta{t}=t_{\mathrm{i}}-t_{\mathrm{i-1}}
\end{equation}
where $t_{i}$ is the burst start time of the $i-$th burst. Due to the unknown waiting time for the first burst in each observation window, only 171 out of a total of 329 bursts could be analyzed (see Fig. \ref{fig:ObsWindow}). A one-sample Kolmogorov-Smirnov test is performed to examine whether the waiting times follow an exponential distribution. The test yields a distance of $D = 0.28$ and $p=0.03$, indicating that a non-homogeneous model is necessary to describe the temporal behavior of the bursts. Therefore, we use the Weibull function to model the waiting times of the non-stationary Poisson process. Specifically, for an observation window of a duration $T$ which owes a number of $N$ bursts, the likelihood of this window can be calculated by \citep{Oppermann2018}, 
\begin{equation}
\begin{aligned}\mathcal{L}(N,t_1,\ldots,t_{\mathrm{N}}|k,\lambda)&=\lambda \operatorname{F}(t_1|k,\lambda)\operatorname{F}(T-t_{\mathrm{N}}|k,\lambda)\\&\times\prod_{i=1}^{N-1}\mathcal{W}(t_{\mathrm{i}+1}-t_{\mathrm{i}}|k,\lambda)\end{aligned}
\end{equation}
with
\begin{equation}
\mathrm{F}(\Delta{t}|k,\lambda)=\mathrm{e}^{-[\Delta{t} \lambda\Gamma(1+1/k)]^k}
\end{equation}
where $\Gamma$ is the Gamma function and $\mathcal{W}(\Delta{t} \mid k, \lambda)$ is the Weibull function. Two free parameters $k$ and $\lambda$ are involved, which correspond to the shape of the Weibull function and the mean burst rate, respectively. The meaning of the Weibull function represents the probability density of the waiting time $\Delta t$, which can be expressed as
\begin{equation}
\mathcal{W}(\Delta{t} \mid k, \lambda) \quad=k \Delta{t}^{-1}[\Delta{t} \lambda \Gamma(1+1 / k)]^{k} e^{-[\Delta{t} \lambda \Gamma(1+1 / k)]^{k}}
\end{equation}

When we use the above expressions to describe a series of burst events, we would conclude that the burst events can be clustered in the timeline if a value of $k<1$ is obtained by maximizing the sum of the likelihoods of all observation windows. Alternatively, if a value of $k=1$ is obtained, then it means the Weibull distribution has reduced to the Poissonian case. By making such a calculation, we finally obtain the corner diagram for the $k-\lambda$ parameter space as shown in Fig. \ref{fig:mcmc}, which is derived from a Markov Chain Monte Carlo (MCMC) survey in the $k-\lambda$ space with 100,000 steps. To be specific, the parameters are constrained to be $\lambda=58.04_{-4.05}^{+4.28}$ day$^{-1}$ and $k=0.70 \pm 0.03$, which demonstrates that the bursts indeed have time clustering and do not occur randomly.

\begin{figure*}
\centering
\centerline{\includegraphics[width=0.75\textwidth]{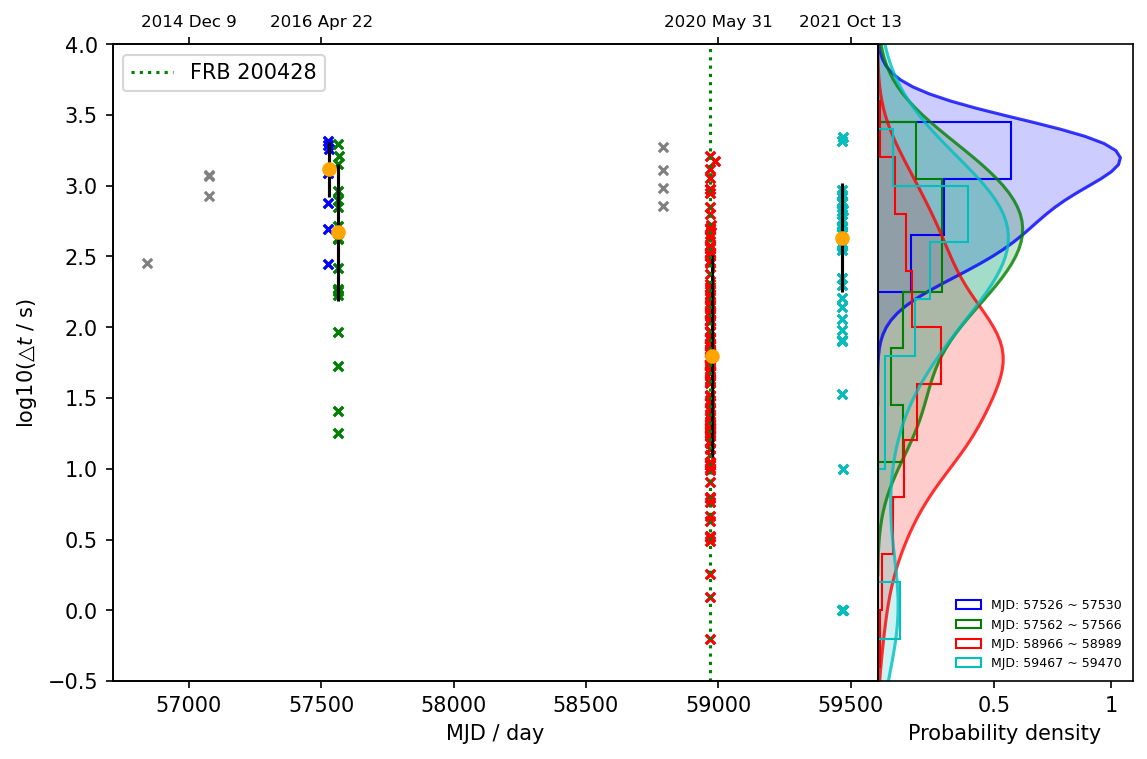}}
\caption{The logarithmic waiting times, $\log(\Delta{t})$, of the XRBs v.s their occurring times, where the time of FRB 20200428 is represented by the green dashed line. The number distribution of the bursts on the waiting times are exhibited in the right panel with the histograms, while the smooth lines give the kernel density of these distributions. The histogram distributions can be approximately fitted by a Gaussian function and its mean value and variance are represented by the orange dot and the short black line in the left panel. 
}\label{fig:WaitMJD}
\end{figure*}

\subsection{The cumulative distribution of the waiting times} \label{subsec:cd}

In Fig. \ref{fig:WaitMJD}, we plot the waiting times of the XRBs of SGR J1935+2154 v.s their burst times, where an apparent clustering of bursts appear. Therefore, we can divide these bursts into 4 separated active episodes as labelled by different colors, except several bursts (grey data) outside of these episodes. Here, an active episode can be defined as a period during which more than two bursts are produced within 10 days, as previously adopted in \cite{Lin2020a,Lin2020b}. The starting and ending times of the four active episodes are listed in Table \ref{tab:wt_log_gaus}. According to the shape of the kernel of the histogram distributions of $\log_{10}(\Delta{t})$ shown in the right panel of Fig. \ref{fig:WaitMJD}, we tentatively use a Gaussian function to fit these histogram distributions. The obtained values of the mean and variance of the Gaussians are listed in Table \ref{tab:wt_log_gaus}. It can be found that the mean value of the waiting times in the 3rd episode (Apr.-May 2020) during which FRB 200428 occurred is significantly shorter than those in the other episodes, indicating the particularity of the FRB episode. In addition, these near-Gaussian distributions of the waiting times make these burst activities of SGR J1935+2154 similar to the previous burst activities of magnetars \citep{Cheng1996,Gogus1999,Gogus2000,Younes2020,Cai2022a}.

\begin{table*}
\centering
\caption{Parameters for the Gaussian fits of the histogram distribution of $\log_{10}( \Delta{t} / s )$}
\label{tab:wt_log_gaus}
\begin{tabular}{cccccc}
\hline
Episodes &UTC & MJD &  $\mu$ & $\sigma$  \\  % & Dependent & Independent & Dependent & Independent \\
\hline
1&2016-05-18 - 2016-05-22      & 57526 - 57530 & 3.11 $\pm$ 0.07 & 0.20 $\pm$ 0.09  \\  % &   7 & 0  & 7  & 0  & 7   \\
2&2016-06-23 - 2016-06-27      & 57562 - 57566 & 2.67 $\pm$ 0.07 & 0.49 $\pm$ 0.07  \\  % &  18 & 1  & 17 & 2  & 16  \\
3&2020-04-27 - 2020-05-20      & 58966 - 58989 & 1.80 $\pm$ 0.06 & 0.71 $\pm$ 0.06  \\  % & 107 & 39 & 68 & 35 & 72  \\
4&2021-09-10 - 2021-09-13      & 59467 - 59470 & 2.63 $\pm$ 0.06 & 0.38 $\pm$ 0.06  \\  % &  31 & 3  & 28 & 5  & 26  \\
\hline
\end{tabular}
\end{table*}

\begin{table*}
\centering
\caption{Parameters for the fittings of the CCD of burst fluence/flux}
\label{tab:pl_best_fit}
\begin{tabular}{llllll}
\hline
         & Model   & $\alpha_1$        & $\alpha_2$        & $\delta$         & $S_{\rm b}$ (or $F_{\rm b}$)  \\
\hline
P( $>S$ ) & BPL & -0.02 $\pm$ 0.01  &  0.66 $\pm$ 0.06  &  0.61 $\pm$ 0.04 & 6.25 $\pm$ 0.73 $\times$ $10^{-8}$ erg cm$^{-2}$ \\
          & SPL &                   &  0.69 $\pm$ 0.01  &                  &   \\
\hline
P( $>F$ ) & BPL & -0.16 $\pm$ 0.07  &  0.92 $\pm$ 0.19  &  1.04 $\pm$ 0.14 & 6.87 $\pm$ 0.31 $\times$ $10^{-7}$ erg cm$^{-2}$ s$^{-1}$ \\
          & SPL &                   &  0.95 $\pm$ 0.02  &                  &       \\
\hline
\end{tabular}
\end{table*}

\begin{figure*}
\centering
\centerline{\includegraphics[width=0.5\textwidth]{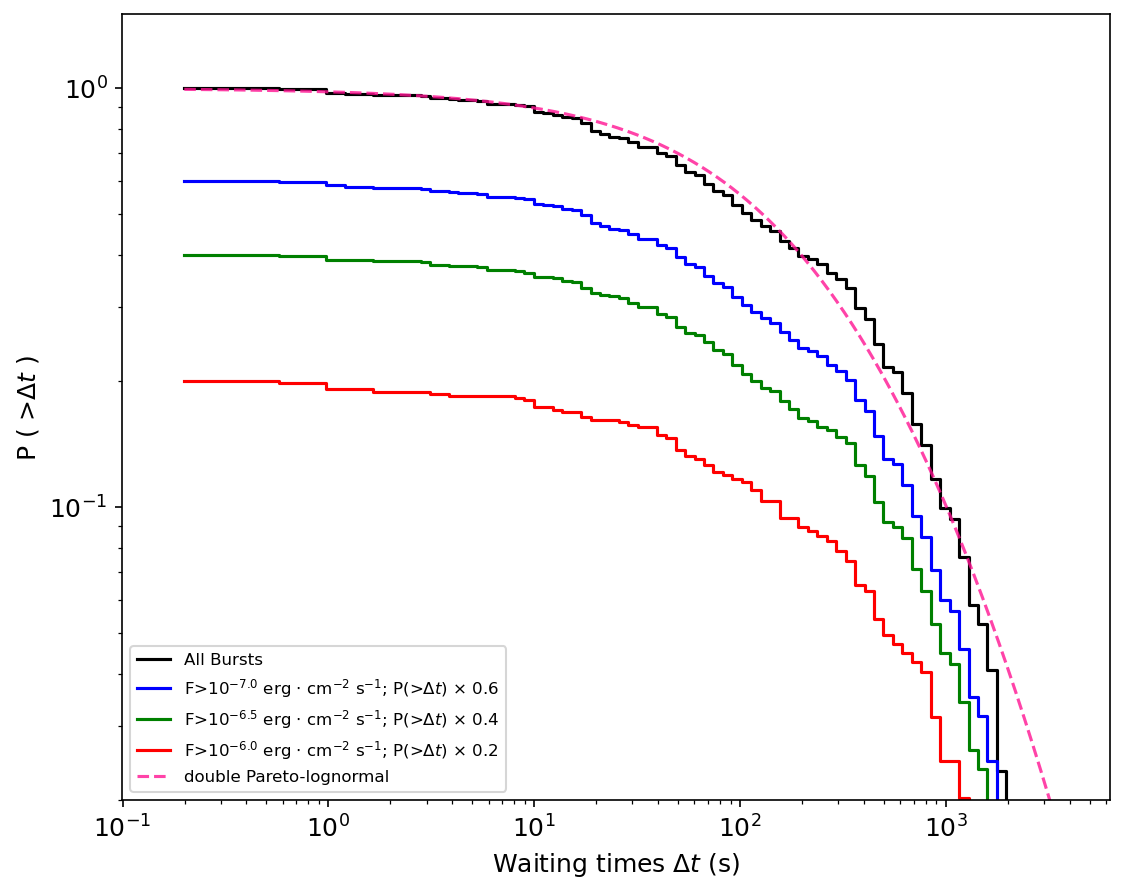}\includegraphics[width=0.5\textwidth]{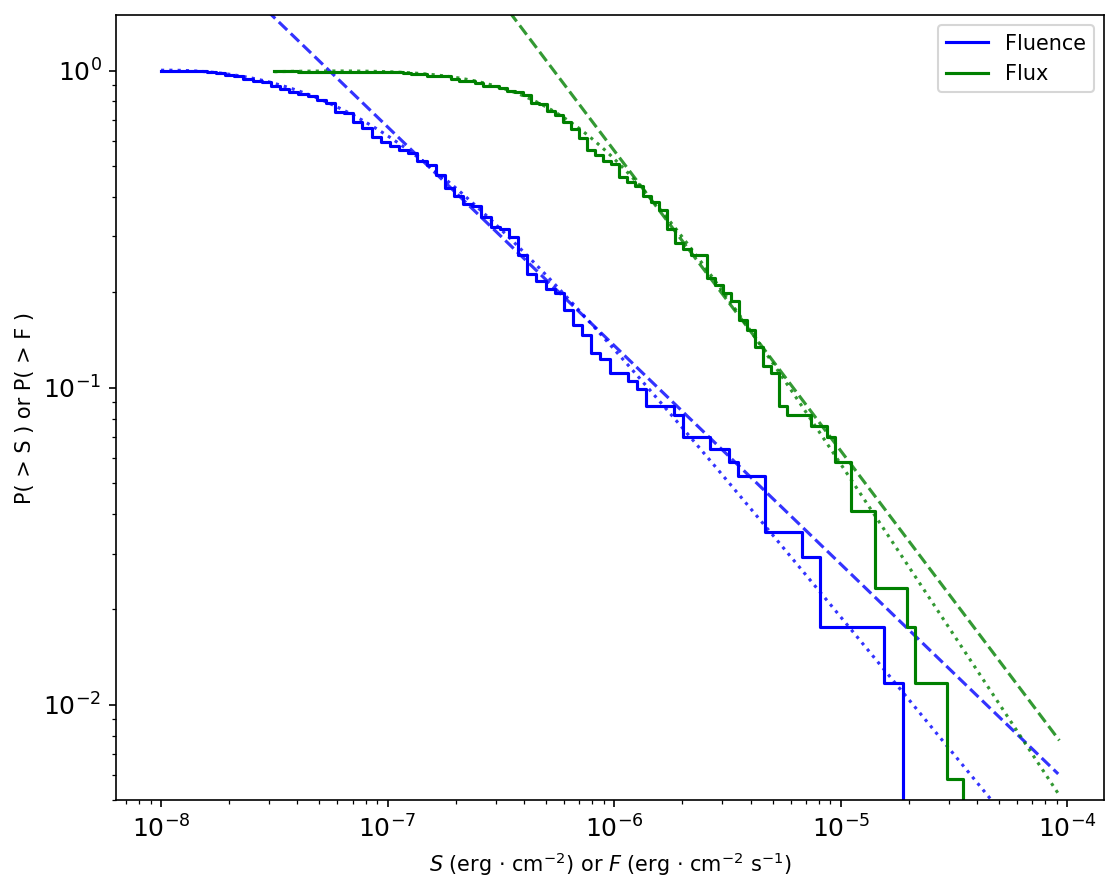}}
\caption{The CCDs of the waiting time (Left) and fluence/flux (Right) of the XRBs. In the left panel, the dashed line gives the dPLn fitting of the CCD, while the dot and dashed lines in the right panel present the smoothly BPL and SPL fits, respectively. }\label{fig:Dis_WT_Flu}
\end{figure*}

For a more detailed description for the distribution property of the waiting times, we plot their complementary cumulative distribution (CCD)\footnote{The cumulative distribution (CD) and CCD of the variable $x$ are defined in this paper as follows:
\begin{equation}
\begin{aligned}
CD(x)&=P(\leq x) = \frac{ n(\leq x) }{N} \\ 
CCD(x)&=P(>x)=1-CD(x)
\end{aligned}
\end{equation}
where $N$ is the total number of the events. }, $P(>\Delta{t})$, in the left panel of Fig. \ref{fig:Dis_WT_Flu}, where the distributions for different flux ranges are also displayed for comparison. According to the heavy-tail feature of these CCDs and  following \cite{Reed2004TheDP}, we try to use the double Pareto-Lognormal (dPLn) distribution to describe their general properties. The dPLn distribution, which consists of two Pareto tail distributions with a transition point in between, arises from a geometric Brownian motion process, that will be killed when the killing rate comes to constant and the state becomes equilibrium \citep{Mitzenmacher2004ABH}. Specifically, the dPLn description for the CCD of the waiting times can be expressed as  \citep{Reed2004TheDP}
\begin{eqnarray}
P(\geq \Delta{t}) &=& \Phi^c\Big(\frac{\log (\Delta{t})-\nu}\tau\Big) +\frac{1}{\alpha_1+\alpha_2} \Big[\alpha_2 \Delta{t}^{-\alpha_1} \nonumber \\
& & A(\alpha_1,\nu,\tau)\Phi\Big(\frac{\log (\Delta{t})-\nu-\alpha_1\tau^2}\tau\Big) +\alpha_1 \Delta{t}^{\alpha_2} \nonumber \\
& & A(-\alpha_2,\nu,\tau)\Phi^c\Big(\frac{\log (\Delta{t})-\nu+\alpha_2\tau^2}\tau\Big)\Big],
\end{eqnarray}
where $\Phi$/$\Phi^c$ is the CD/CCD of standard normal distribution, $\nu$ and $\tau^2$ represent the mean and variance of the log-normal for the initial geometric Brownian motion process, and
\begin{equation}
A(\alpha,\nu,\tau)=\exp(\alpha\nu+\alpha^2\tau^2/2).
\end{equation}
The parameters $\alpha_1$ and $\alpha_2$ determine the slope of the two Pareto tails. By using the package \textit{distribuisonsrd} of Comprehensive R Archive Network \citep[CRAN,][]{MarçalSalvadorElena2020}, we perform the dPLn fitting to the CCDs of the waiting times, as shown by the dashed line in Figure \ref{fig:Dis_WT_Flu}, and obtain $\alpha_1$=3.85 and $\alpha_2$=0.73. This indicates that, in the range of a relatively short waiting time, the CCD can be approximated by a power law with a slope of $\alpha_2 \sim 0.73$, which could further be independent of the selection of the flux threshold and thus of the burst energy.

\subsection{The cumulative distribution of the fluence/flux}

In the right panel of Fig. \ref{fig:Dis_WT_Flu}, we further show the CCD of the fluence of the XRBs, which can be fitted by a smoothly broken-power law (BPL) as
\begin{equation} 
P( > S ) \propto {\left(\frac{S}{S_{b}}\right)^{-\alpha_1}}\left\{\frac12\left[1+\left(\frac{S}{S_{b}}\right)^{1/\delta}\right]\right\}^{(\alpha_1-\alpha_2)\delta}
\label{eq:smooth_pl}
\end{equation}
with the parameter values listed in Table \ref{tab:pl_best_fit}. According to the above-obtained value of $S_b$ (or $F_b$), we further ignore the CCD in the low fluence/flux range and fit the high-fluence/flux distribution by a single power law (SPL), which are also shown in the right panel of Fig. \ref{fig:Dis_WT_Flu} by the dashed lines. This result indicates that the CCD of the burst fluence/flux can basically be described by a SPL, which is consistent with the results reported in previous literature \citep{vanderHorst2012,Collazzi2015,Lin2020a}.

%%%%%%%%%%%%%%%%%%%%%%%%%%%%%%%%%%%%%%%%%%%%%
\section{Unified scaling law for magnetar bursts} \label{sec:usl}
The temporal clustering of the XRBs of SGR J1935+2154 and the distributions of their waiting time and fluence/flux make these burst activity very similar to the earthquake phenomena. The earthquakes also satisfy two power-law frequency–size statistics, i.e., the \textit{Omori law} and \textit{Gutenberg-Richter law} as
\begin{equation}
p(\Delta{t}) \propto \Delta{t}^{-\alpha_{\rm E}}
\end{equation}
with $\alpha_{\rm E}\sim (0.34-1.83)$ and
\begin{equation}
P(>S_{\mathrm{E}})\propto S_{\mathrm{E}}^{-\beta_{\rm E}}
\end{equation} 
with $\beta_{\rm E}\sim (0.6-1.0)$, where $S_{\mathrm{E}}$ is the maximum seismic wave amplitude which determines the magnitude of the aftershocks of the earthquake as $m = \mathrm{log}_{10} (S_{\mathrm{E}})$ \citep{Aschwanden2011book}. Earthquakes are known to be SOC phenomena. Under the long-term action of external forces, such a SOC system can always enter into a statistically stationary state characterized by spatial and temporal scale independence, without fine-tuning the system parameters. In this state, a minor change can lead to an avalanche and a collapse of the whole system. Therefore, it can be naturally understood why the aftershocks of earthquakes can have temporal clustering and power-law features of CD. Moreover, the connection between these two characteristics can in principle be understood according to the analysis by \cite{AschwandenMarkus2010}, who demonstrated that the waiting time can be distributed as
\begin{eqnarray} 
p(\Delta t)&=&\frac{\lambda}{(1+\lambda\Delta t)^2}\nonumber\\
&\approx&(1/\lambda)\Delta t^{-2}, {\rm ~for~ }\Delta{t} \gg 1/\lambda,
\label{eq:cluster}
\end{eqnarray} 
if the variability of the event rate shows a spike-like $\delta$-function with $\lambda$ being the mean event rate. For the burst activity of SGR J1935+2154, we have $\lambda=58.04_{-4.05}^{+4.28}$ day$^{-1}$ as obtained in Section \ref{sec:tp}. 

\begin{figure*}
\centering
\centerline{\includegraphics[width=0.5\textwidth]{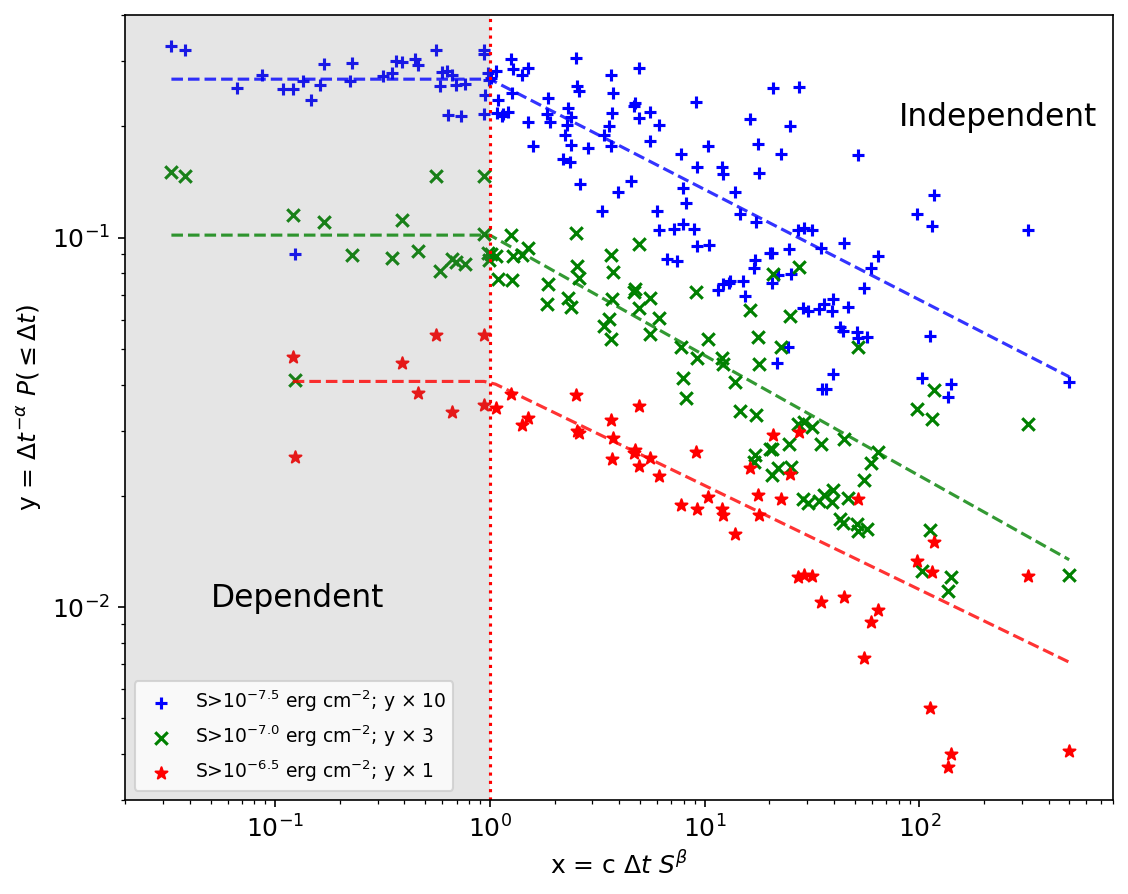}\includegraphics[width=0.5\textwidth]{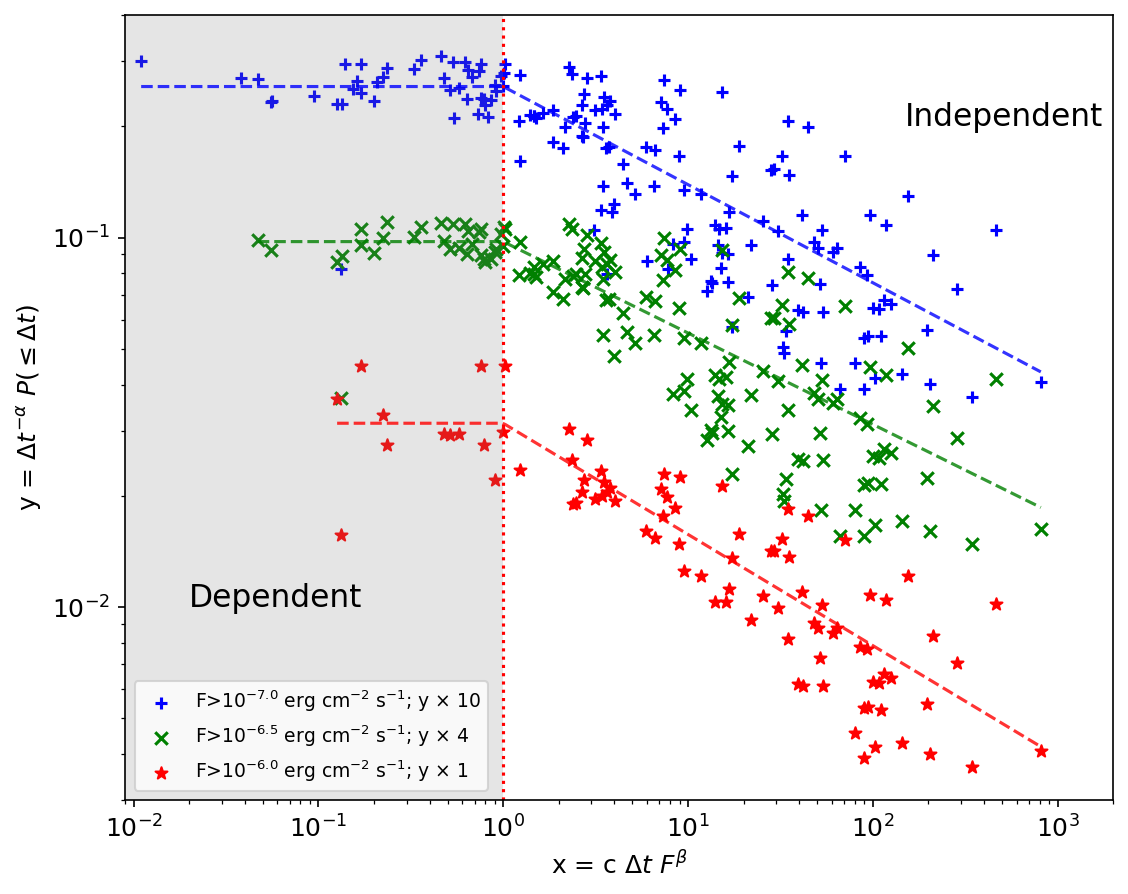}}
\caption{The $x-y$ USL of XRBs. The dashed lines give the fits of the data by Eq. (\ref{eq:usl_bpl}) for parameter values listed in Table \ref{tab:usl}.}\label{fig:USL}
\end{figure*}

As a SOC system, the earthquake phenomena had further been found to satisfy a unified scaling law (USL) between the quantities of $\Delta{t} S_{\mathrm{E}}^{-\beta_{\rm E}} L^{d_f}$ and $\Delta{t}^{-\alpha_{\rm E}} p(\Delta{t})_{S,L}$, where $L$ is the fault lengths, $d_f$ is the fractal dimension, and $p(\Delta{t})_{S,L}$ is the distribution of waiting times $\Delta{t}$ between earthquakes occurring within range $L$ whose amplitude greater than $S_{\mathrm{E}}$ \citep{Bak2002}. Here, the scaling variable $\Delta{t} S^{-\beta_{\rm E}} L^{d_f}$ can be interpreted as a measure of the average number of events occurring within a time interval $\Delta{t}$ and a range $L$ with an amplitude greater than $S_{\mathrm{E}}$. Then, in view of the similarity between the earthquakes and the magnetar burst activities, it is supposed that, in principle, a similar USL could also be found from the XRBs of magnetars, which reads 
\begin{equation}\label{eq:usl_bpl}
y=f(x)\approx\left\{
\begin{array}{cc}
{\rm const.},     & {\rm for ~}x<1 ,\\
x^{-\gamma},     & {\rm for ~}x\geq1 ,
\end{array} 
\right.
\end{equation} 
where $x \equiv c\Delta{t}S^{\beta}$ and $y\equiv \Delta{t}^{-\alpha}P(\leq \Delta{t})$, $c$ is coefficient, and $S$ represents fluence which can be replaced with $F$ if flux is used. The scaling variable $\Delta{t}S^{\beta}$ (or $\Delta{t}F^{\beta}$) can be interpreted as a measure of the average number of bursts occurring within a time interval $\Delta{t}$ with a fluence/flux less than $S$ (or $F$). By using the parameter values of $\alpha=0.73$ derived from the CCD of waiting times and $\beta=0.69$ (or 0.95) from the CCD of fluence/flux, we plot the distribution of the XRBs of SGR J1935+2154 in the $x-y$ plane in Fig. \ref{fig:USL}, where an appropriate value $c$ is adopted to define $x$ in order to set the break value of $x$ at $1$. As shown, the distribution of the observational data indeed all behave as described in Eq. (\ref{eq:usl_bpl}), which have a constant-$y$ and decaying-$y$ part with parameters listed in Table \ref{tab:usl}. This result is insensitive to the selection of the fluence/flux limitation.

\begin{table}
\centering
\caption{Parameters for the USL presented in Eq. (\ref{eq:usl_bpl})}
\label{tab:usl}
\begin{tabular}{cc|cc}
\hline
&Selection       & $c/10^{3}$     & $\gamma$           \\
\hline
        &$ > 10^{-7  }$ &    & -0.30 $\pm$ 0.02   \\
Fluence $S$   &$ > 10^{-6.5}$ &    1.68    & -0.33 $\pm$ 0.02   \\
 (erg cm$^{-2}$)      &$ > 10^{-6  }$ &        & -0.28 $\pm$ 0.02   \\
\hline
        &$ > 10^{-7  }$ &   & -0.27 $\pm$ 0.02   \\
Flux $F$ &$ > 10^{-6.5}$ &  19.67      & -0.24 $\pm$ 0.01   \\
(erg cm$^{-2}$ s$^{-1}$)       &$ > 10^{-6  }$ &        & -0.30 $\pm$ 0.02   \\
\hline

\end{tabular}
\end{table}

\begin{figure*}
\centering
\centerline{\includegraphics[width=0.5\textwidth]{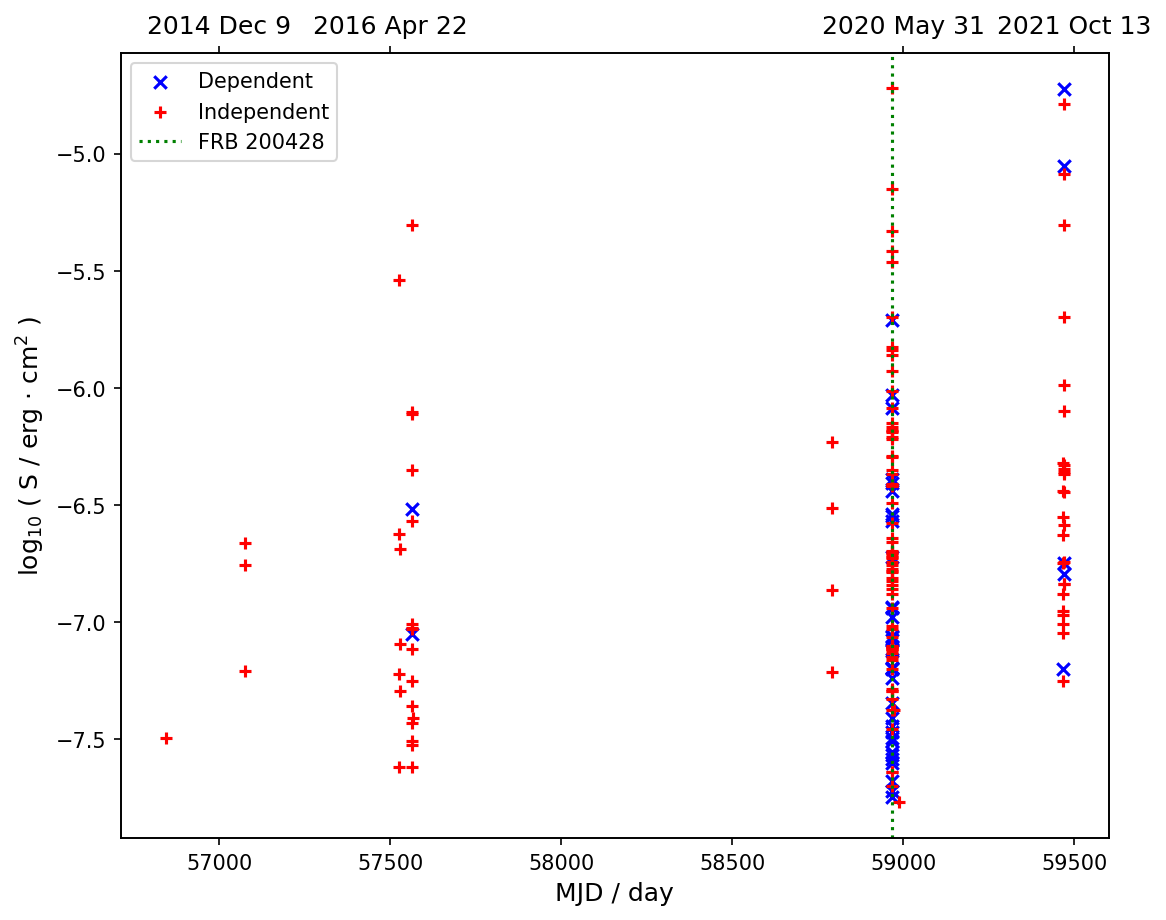}\includegraphics[width=0.5\textwidth]{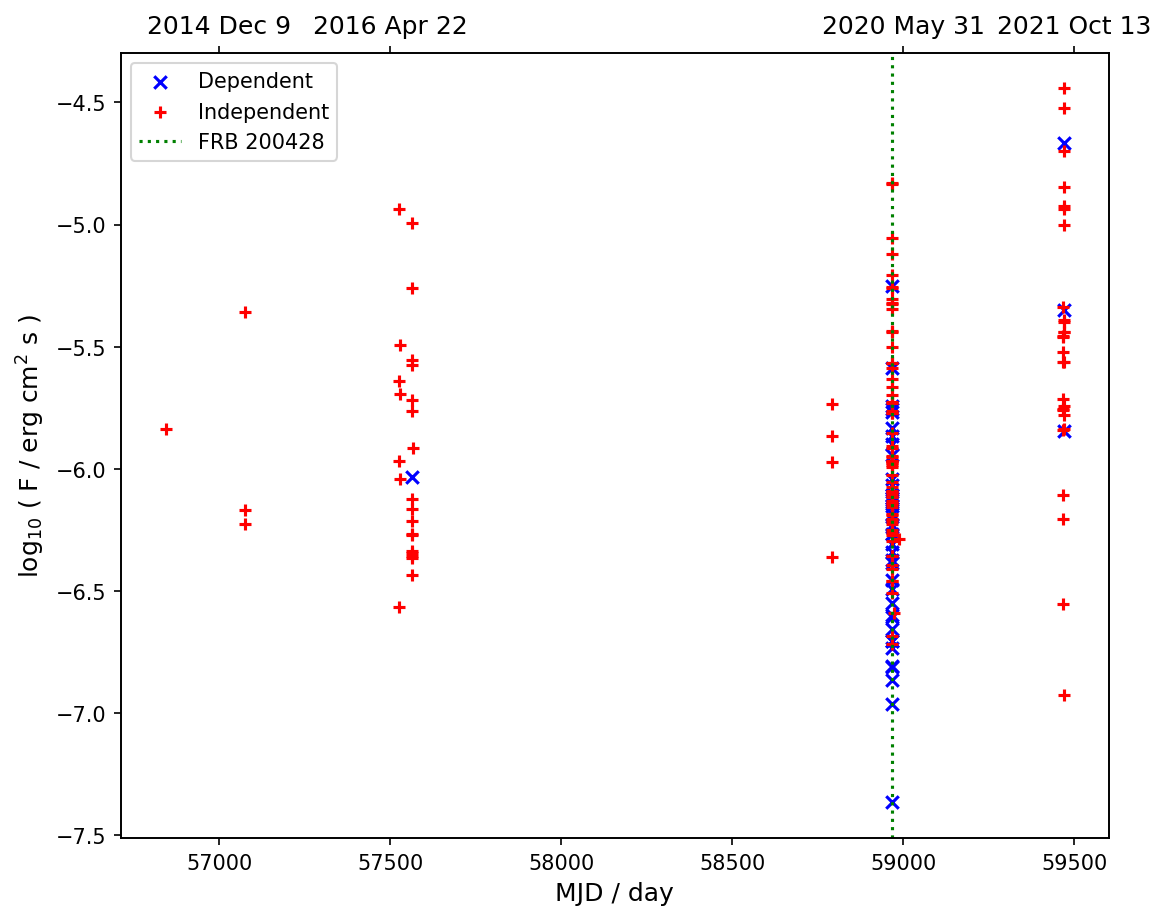}}
\caption{The fluence/flux vs. the occurring time of XRBs, where the bursts have been separated into the dependent and independent types by according the USL distribution presented in Fig. \ref{fig:USL}.}\label{fig:FluMJD}
\end{figure*}

\begin{figure}
\centering
\centerline{\includegraphics[width=0.5\textwidth]{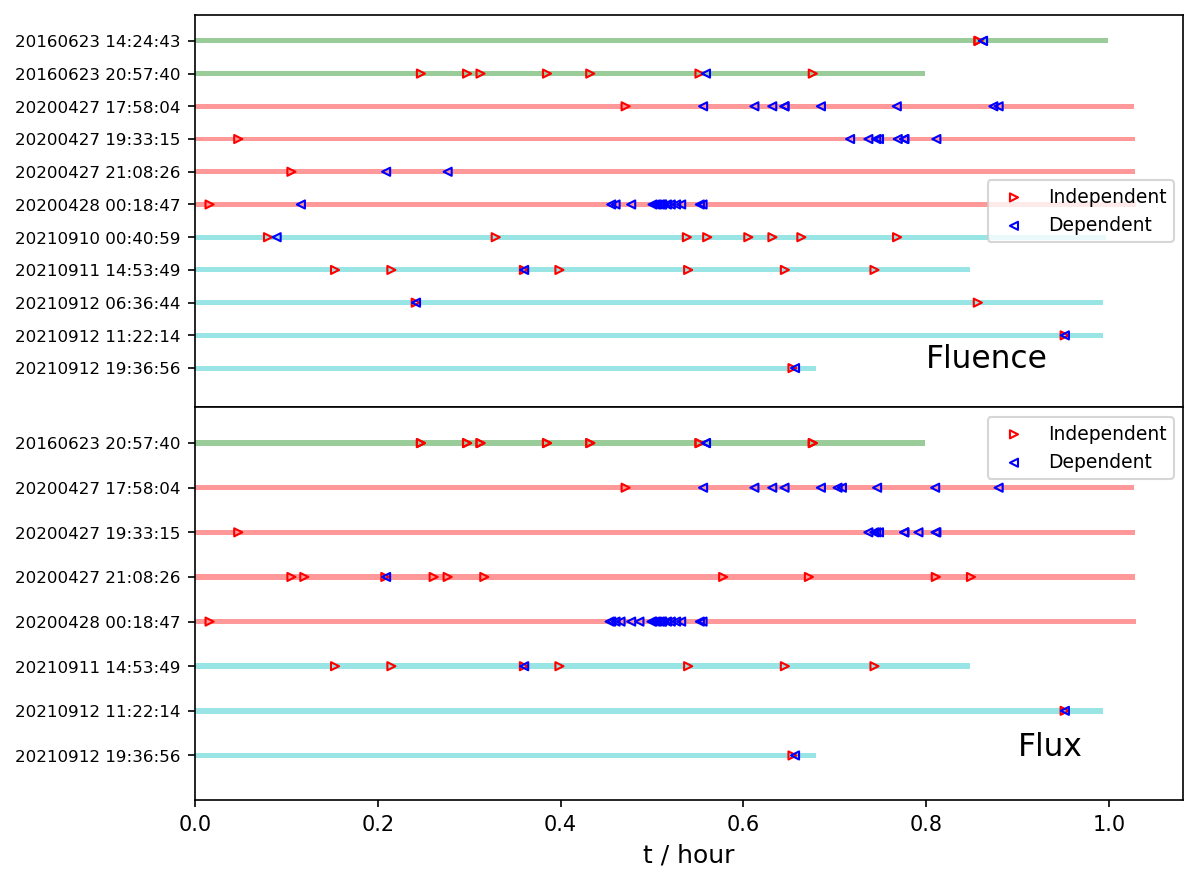}}
\caption{The observation windows owing dependent bursts. The upper and lower panels are plotted corresponding to results presented in the left and right panels of Fig. \ref{fig:FluMJD}, respectively. 
}\label{fig:ObsWin_S}
\end{figure}

In more detail, the USL indicates that the bursts locating the constant part of $f(x)$, which corresponds to the \textit{Omori} law in the earthquake case, are likely to be dependent afterbursts of some main bursts. Meanwhile, the events deviating from the \textit{Omori} law can be classified as independent burst events. This classification can be basically in agreement with the physical intuition that XRBs that are separated by large energy difference or long waiting times could be independent. According to this statistical classification, we plot the inferred independent and dependent bursts in Figs. \ref{fig:FluMJD} and \ref{fig:ObsWin_S} by using different symbols, the specific properties of which are listed in Tables \ref{tab:depend_burst_flux}-\ref{tab:independ_burst_fluence}. In Table \ref{tab: de-in-num}, we list the numbers of the dependent and independent bursts for the four different episodes. It can be found that the number fraction of dependent bursts in the total bursts in the FRB episode (Apr.-May 2020) is overwhelmingly higher than those in the other three episodes. Specifically, the faction of the dependent bursts (i.e., the number fraction of the data of $x<1$) can be calculated to be $\sim$33\% (or  $\sim$36\%) for the fluence (or flux) definition of $x$. The observation windows owing these dependent bursts are plotted in Fig. \ref{fig:ObsWin_S}. According to this result, it can be naturally understood why the FRB episode has the smallest mean value of the waiting times, as found from Fig. \ref{fig:WaitMJD}. The large fraction of the dependent bursts increase the density of bursts in timeline and thus shorten the apparent waiting time between the successive bursts. Following this consideration, it is suspected that  FRBs or slow radio bursts could exist in the last episode (Sept. 2021), because several dependent bursts appear there.

In addition, it needs to be noticed that, due to the limited detection efficiency of the instrument and the criteria used to distinguish the bursts, a bright long burst may contain an independent burst and multiple dependent bursts. The number of dependent bursts is likely underestimated. The energy of a burst may be the sum of multiple dependent ones and may be misclassified as independent (see left panel of Fig. \ref{fig:FluMJD}). Therefore, using the peak flux is recommended for statistical analysis of magnetar bursts.

\begin{table*}
\centering
\caption{Numbers of the inferred dependent/independent bursts}
\label{tab: de-in-num}
\begin{tabular}{c|c|cc|cc}
\hline
Episodes & Total & \multicolumn2c{Classified with flux} & \multicolumn2c{Classified with fluence} \\
\cline{3-6}%\cline{6-9}
&& Dependent & Independent & Dependent & Independent \\
\hline
1 & 7 & 0  & 7  & 0  & 7   \\
2 & 18 & 1  & 17 & 2  & 16  \\
3 & 107 & 39 & 68 & 35 & 72  \\
4 & 31 & 3  & 28 & 5  & 26  \\
\hline
\end{tabular}
\end{table*}

%%%%%%%%%%%%%%%%%%%%%%%%%%%%%%%%%%%%%%%%%%%%%
\section{Conclusion and discussion} \label{sec:discus}

The distribution of the waiting time and fluence/flux of the XRBs of SGR J1935+2154 are investigated statistically in this paper, which firstly show the temporal clustering of the bursts and the power-law behavior of the distributions. These results indicate that the XRBs can belong to a SOC phenomenon, which could correspond to sudden fractures and displacements of the crust of the magnetar \citep[i.e.,  starquakes;][]{ThompsonDuncan1995,Cheng1996}. Our statistics further show that, different from the other episodes, the episode during which FRB 20200428 occurred has an obviously smaller mean value of the waiting time, which directly indicates the particularity of this episode. In more detail, by applying the USL discovered by \cite{Bak2002} for earthquakes, it is found that the regular episodes could mainly consist of independent bursts, where as much more dependent bursts can happen in the FRB episode. Nevertheless, it should be noted that we actually cannot 
exactly specify the individual bursts to be dependent or independent, only based on the statistical investigation.

In any case, our statistical results indicate that the starquakes occurring in regular episodes could be generally without causal link (i.e., independent). On the contrary, in the FRB episode, independent starquakes could take place more frequently and intensively and, thus, many secondary bursts could be induced by the interactions among these primary independent bursts. This further hints that FRB emission could result from such burst interactions (e.g., the collision between different seismic/Alfven waves or different explosion outflows) and be associated with a dependent burst. Such a situation could appear in a global activity of a magnetar due to a conversion of the magnetic field from a dipolar to a multi-polar configuration \citep[e.g., the situation considered in][]{Yang2021ApJ}. Here, it is interesting to mention that a giant glitch occurred approximately 3 days before FRB 20200428, as reported recently by \cite{Ge2022arxiv}. The occurrence of the glitch means the magnetar crust had just obtained a sudden increase of the angular momentum, which could be transferred from the core of the magnetar. In such a case, due to the increase of the differential rotation between the crust and the core, the stress forced on the whole crust by the internal magnetic field could be consequently strengthened to exceed the elastic yield limit. Therefore, it can be imagined that crust cracking events could take place nearly globally and, simultaneously, a multi-polar field can be formed.

After the FRB 20200428 event, it has further been reported that the Gravitational Wave High-energy Electromagnetic Counterpart All-sky Monitor (GECAM) detected an XRB from SGR J1935+2154 at an arrival time consistent with a radio burst detected by CHIME \citep{Wang2022ATel} on 2020 October 14 UTC 19:21:39.100. The Insight–Hard X-ray Modulation Telescope (HXMT) also detected an XRB associated with a radio burst detected by the Yunnan 40 m telescope, on 2022 October 21 \citep{Li2022ATel}. Future works on these more XRB-FRB association events would help us to test the discovery in this paper.

% Acknowledgments
% \begin{acknowledgments}
This work is supported by the National Key R\&D Program of China (2021YFA0718500), the National SKA program of China (2020SKA0120300), the National Natural Science Foundation of China (Grant No. 12393811, 12173038), and the Strategic Priority Research Program on Space Science (Grant No. XDA15360102, XDA15360300, XDA15052700) of the Chinese Academy of Sciences.
% \end{acknowledgments}

% Appendix
% \clearpage
\appendix
\section{The X-ray bursts classification of SGR J1935+2154} \label{sec:appendix_a}

The inferred dependent/independent bursts classified by flux/fluence are listed here. The Flux in units of $\times 10^{-7}$ erg $\cdot$ cm$^{-2}$ s$^{-1}$ and Fluence in units of $\times 10^{-7}$ erg $\cdot$ cm$^{-2}$. The time range of the 1st, 2nd, 3rd and 4th episodes is May 2016, June 2016, Apr.-May 2020 and Sept. 2021 respectively (see Table \ref{tab:wt_log_gaus}). Tables \ref{tab:depend_burst_flux}-\ref{tab:independ_burst_fluence} present the inferred dependent/independent bursts classified by flux/fluence.

\clearpage
\begin{table}
\scriptsize
% \centering
\caption{The list of the inferred dependent bursts classified by flux}
\label{tab:depend_burst_flux}
\begin{tabular}{lccc}
\hline
Episode & Time (UTC) & Flux & Fluence \\
\hline
2nd episode & 2016-06-23T21:31:12.600 & 9.27 & 0.89 \\
3rd episode & 2020-04-27T18:31:25.234 & 11.39 & 1.89 \\
            & 2020-04-27T18:34:47.296 & 1.97 & 1.05 \\
            & 2020-04-27T18:35:57.633 & 18.00 & 0.45 \\
            & 2020-04-27T18:36:46.007 & 56.13 & 19.42 \\
            & 2020-04-27T18:39:09.331 & 5.43 & 0.19 \\
            & 2020-04-27T18:40:15.043 & 2.52 & 1.15 \\
            & 2020-04-27T18:40:32.031 & 4.55 & 6.16 \\
            & 2020-04-27T18:42:50.652 & 25.85 & 8.17 \\
            & 2020-04-27T18:46:39.414 & 5.88 & 9.70 \\
            & 2020-04-27T18:50:49.460 & 7.43 & 0.26 \\
            & 2020-04-27T20:17:27.317 & 9.06 & 0.58 \\
            & 2020-04-27T20:17:50.343 & 1.55 & 2.20 \\
            & 2020-04-27T20:17:58.442 & 14.74 & 1.15 \\
            & 2020-04-27T20:18:09.130 & 0.43 & 0.36 \\
            & 2020-04-27T20:19:47.631 & 1.37 & 1.16 \\
            & 2020-04-27T20:19:49.430 & 17.63 & 4.09 \\
            & 2020-04-27T20:20:44.640 & 2.21 & 0.74 \\
            & 2020-04-27T20:21:51.841 & 2.82 & 1.63 \\
            & 2020-04-27T20:21:55.136 & 17.01 & 9.34 \\
            & 2020-04-27T21:20:58.670 & 4.12 & 0.77 \\
            & 2020-04-28T00:46:00.034 & 7.58 & 6.05 \\
            & 2020-04-28T00:46:06.394 & 4.94 & 0.87 \\
            & 2020-04-28T00:46:23.528 & 3.21 & 2.71 \\
            & 2020-04-28T00:46:43.072 & 7.59 & 3.82 \\
            & 2020-04-28T00:47:24.957 & 3.51 & 0.84 \\
            & 2020-04-28T00:47:57.536 & 5.77 & 1.50 \\
            & 2020-04-28T00:48:44.833 & 4.24 & 1.88 \\
            & 2020-04-28T00:48:49.098 & 13.69 & 8.17 \\
            & 2020-04-28T00:49:00.270 & 1.09 & 2.85 \\
            & 2020-04-28T00:49:06.479 & 7.78 & 0.21 \\
            & 2020-04-28T00:49:16.609 & 12.59 & 3.94 \\
            & 2020-04-28T00:49:22.392 & 7.03 & 0.64 \\
            & 2020-04-28T00:49:27.008 & 1.84 & 0.64 \\
            & 2020-04-28T00:49:45.895 & 2.47 & 2.88 \\
            & 2020-04-28T00:50:01.012 & 7.29 & 3.64 \\
            & 2020-04-28T00:50:21.993 & 8.57 & 0.18 \\
            & 2020-04-28T00:50:41.835 & 1.56 & 0.63 \\
            & 2020-04-28T00:51:55.444 & 6.52 & 0.86 \\
            & 2020-04-28T00:52:06.141 & 1.97 & 0.75 \\
            & 2021-09-11T15:15:26.000 & 216.73 & 188.99 \\
            & 2021-09-12T12:19:21.000 & 44.75 & 1.79 \\
4th episode  & 2021-09-12T20:16:20.000 & 14.29 & 1.60 \\
\hline
\end{tabular}
% \tablecomments{The Flux in units of $\times 10^{7}$ erg $\cdot$ cm$^{-2}$ s$^{-1}$ and Fluence in units of $\times 10^{7}$ erg $\cdot$ cm$^{-2}$. The time range of the 1st, 2nd, 3rd and 4th episodes is May 2016, June 2016, Apr.-May 2020 and Sept. 2021 respectively (see Table \ref{tab:wt_log_gaus}). }
\end{table}

\begin{table}
\scriptsize
% \centering
\caption{The list of the inferred dependent bursts classified by fluence}
\label{tab:depend_burst_fluence}
\begin{tabular}{lccc}
\hline
Episode & Time (UTC) & Flux & Fluence \\
\hline
2nd episode & 2016-06-23T15:16:26.836 & 17.27 & 3.04 \\
            & 2016-06-23T21:31:12.600 & 9.27 & 0.89 \\
3rd episode & 2020-04-27T18:31:25.234 & 11.39 & 1.89 \\
            & 2020-04-27T18:34:47.296 & 1.97 & 1.05 \\
            & 2020-04-27T18:35:57.633 & 18.00 & 0.45 \\
            & 2020-04-27T18:36:45.376 & 20.00 & 0.28 \\
            & 2020-04-27T18:36:46.007 & 56.13 & 19.42 \\
            & 2020-04-27T18:39:09.331 & 5.43 & 0.19 \\
            & 2020-04-27T18:44:08.209 & 4.42 & 0.34 \\
            & 2020-04-27T18:50:28.665 & 10.80 & 0.27 \\
            & 2020-04-27T18:50:49.460 & 7.43 & 0.26 \\
            & 2020-04-27T20:16:15.285 & 10.67 & 0.32 \\
            & 2020-04-27T20:17:27.317 & 9.06 & 0.58 \\
            & 2020-04-27T20:17:58.442 & 14.74 & 1.15 \\
            & 2020-04-27T20:18:09.130 & 0.43 & 0.36 \\
            & 2020-04-27T20:19:23.068 & 10.33 & 0.31 \\
            & 2020-04-27T20:19:47.631 & 1.37 & 1.16 \\
            & 2020-04-27T20:19:49.430 & 17.63 & 4.09 \\
            & 2020-04-27T20:21:55.136 & 17.01 & 9.34 \\
            & 2020-04-27T21:20:58.670 & 4.12 & 0.77 \\
            & 2020-04-27T21:25:01.037 & 6.50 & 0.39 \\
            & 2020-04-28T00:25:43.946 & 5.95 & 0.25 \\
            & 2020-04-28T00:46:06.394 & 4.94 & 0.87 \\
            & 2020-04-28T00:46:23.528 & 3.21 & 2.71 \\
            & 2020-04-28T00:47:24.957 & 3.51 & 0.84 \\
            & 2020-04-28T00:48:49.098 & 13.69 & 8.17 \\
            & 2020-04-28T00:49:00.270 & 1.09 & 2.85 \\
            & 2020-04-28T00:49:06.479 & 7.78 & 0.21 \\
            & 2020-04-28T00:49:16.609 & 12.59 & 3.94 \\
            & 2020-04-28T00:49:22.392 & 7.03 & 0.64 \\
            & 2020-04-28T00:49:27.008 & 1.84 & 0.64 \\
            & 2020-04-28T00:49:45.895 & 2.47 & 2.88 \\
            & 2020-04-28T00:50:01.012 & 7.29 & 3.64 \\
            & 2020-04-28T00:50:21.993 & 8.57 & 0.18 \\
            & 2020-04-28T00:50:41.835 & 1.56 & 0.63 \\
            & 2020-04-28T00:51:55.444 & 6.52 & 0.86 \\
            & 2020-04-28T00:52:06.141 & 1.97 & 0.75 \\
4th episode & 2021-09-10T00:46:21.000 & 17.50 & 0.63 \\
            & 2021-09-11T15:15:26.000 & 216.73 & 188.99 \\
            & 2021-09-12T06:51:14.000 & 300.37 & 88.91 \\
            & 2021-09-12T12:19:21.000 & 44.75 & 1.79 \\
            & 2021-09-12T20:16:20.000 & 14.29 & 1.60 \\
\hline
\end{tabular}
% \tablecomments{The Flux in units of $\times 10^{7}$ erg $\cdot$ cm$^{-2}$ s$^{-1}$ and Fluence in units of $\times 10^{7}$ erg $\cdot$ cm$^{-2}$. The time range of the 1st, 2nd, 3rd and 4th episodes is May 2016, June 2016, Apr.-May 2020 and Sept. 2021 respectively (see Table \ref{tab:wt_log_gaus}). }
\end{table}

\clearpage
\startlongtable
\begin{deluxetable}{lccc}
\tabletypesize{\scriptsize}
\tablecaption{The list of the inferred independent bursts classified by flux}
\label{tab:independ_burst_flux}
\tablehead{\colhead{Episode} & \colhead{Time (UTC)} & \colhead{Flux} & \colhead{Fluence}}
\startdata
outside-episodes\tablenotemark{a} & 2014-07-05T09:37:34.484 & 14.55 & 0.32 \\
            & 2015-02-22T19:44:16.881 & 44.00 & 1.76 \\
            & 2015-02-22T19:58:16.658 & 5.96 & 0.62 \\
            & 2015-02-23T05:24:53.876 & 6.78 & 2.17 \\
1st episode & 2016-05-18T09:09:23.800 & 116.29 & 28.84 \\
            & 2016-05-18T10:28:02.817 & 22.79 & 2.37 \\
            & 2016-05-18T15:33:47.010 & 10.71 & 0.60 \\
            & 2016-05-18T17:05:12.813 & 2.73 & 0.24 \\
            & 2016-05-19T11:59:32.632 & 9.11 & 0.51 \\
            & 2016-05-19T12:07:46.594 & 32.03 & 2.05 \\
            & 2016-05-21T21:13:16.242 & 20.25 & 0.81 \\
2nd episode & 2016-06-23T15:16:26.836 & 17.27 & 3.04 \\
            & 2016-06-23T16:49:57.493 & 26.49 & 4.45 \\
            & 2016-06-23T17:47:06.896 & 7.50 & 0.24 \\
            & 2016-06-23T17:55:48.768 & 4.62 & 0.37 \\
            & 2016-06-23T17:58:55.696 & 4.31 & 0.31 \\
            & 2016-06-23T18:22:43.425 & 5.38 & 0.56 \\
            & 2016-06-23T19:36:27.341 & 4.38 & 0.77 \\
            & 2016-06-23T19:37:59.997 & 6.84 & 0.93 \\
            & 2016-06-23T19:51:16.261 & 3.67 & 0.44 \\
            & 2016-06-23T20:06:37.158 & 28.02 & 2.69 \\
            & 2016-06-23T21:15:31.444 & 6.13 & 0.98 \\
            & 2016-06-23T21:16:24.612 & 19.04 & 7.77 \\
            & 2016-06-23T21:20:46.404 & 101.19 & 49.38 \\
            & 2016-06-23T21:23:36.780 & 54.86 & 7.90 \\
            & 2016-06-23T21:30:47.032 & 4.52 & 0.94 \\
            & 2016-06-23T21:38:13.432 & 5.36 & 0.30 \\
            & 2016-06-26T17:50:03.251 & 12.19 & 0.39 \\
outside-episodes\tablenotemark{a} & 2019-11-04T09:17:53.492 & 18.35 & 5.89 \\
            & 2019-11-04T20:13:42.537 & 4.36 & 0.61 \\
            & 2019-11-04T20:29:39.804 & 10.70 & 1.37 \\
            & 2019-11-04T23:48:01.336 & 13.64 & 3.07 \\
3rd episode & 2020-04-27T18:31:05.770 & 3.11 & 0.76 \\
            & 2020-04-27T18:33:53.116 & 7.18 & 0.51 \\
            & 2020-04-27T18:34:05.700 & 47.65 & 20.11 \\
            & 2020-04-27T18:34:46.047 & 16.95 & 3.83 \\
            & 2020-04-27T18:35:05.320 & 49.71 & 5.12 \\
            & 2020-04-27T18:35:46.623 & 21.64 & 1.32 \\
            & 2020-04-27T18:36:45.376 & 20.00 & 0.28 \\
            & 2020-04-27T18:38:20.206 & 8.86 & 0.93 \\
            & 2020-04-27T18:38:53.689 & 7.24 & 1.81 \\
            & 2020-04-27T18:42:40.816 & 11.29 & 0.35 \\
            & 2020-04-27T18:44:08.209 & 4.42 & 0.34 \\
            & 2020-04-27T18:46:08.767 & 18.74 & 3.86 \\
            & 2020-04-27T18:47:05.754 & 76.00 & 11.78 \\
            & 2020-04-27T18:48:38.675 & 6.95 & 1.69 \\
            & 2020-04-27T18:49:28.034 & 18.56 & 6.83 \\
            & 2020-04-27T18:50:28.665 & 10.80 & 0.27 \\
            & 2020-04-27T18:55:44.155 & 36.81 & 2.65 \\
            & 2020-04-27T18:57:35.574 & 8.12 & 0.78 \\
            & 2020-04-27T18:58:45.533 & 4.04 & 0.78 \\
            & 2020-04-27T19:37:39.328 & 5.37 & 3.89 \\
            & 2020-04-27T19:43:44.537 & 88.12 & 38.42 \\
            & 2020-04-27T19:45:00.478 & 5.05 & 0.51 \\
            & 2020-04-27T19:55:32.325 & 5.61 & 0.23 \\
            & 2020-04-27T20:01:45.681 & 10.52 & 5.08 \\
            & 2020-04-27T20:07:20.319 & 2.07 & 0.79 \\
            & 2020-04-27T20:13:38.263 & 9.45 & 0.52 \\
            & 2020-04-27T20:14:51.396 & 27.06 & 1.38 \\
            & 2020-04-27T20:15:20.583 & 148.03 & 189.77 \\
            & 2020-04-27T20:16:15.285 & 10.67 & 0.32 \\
            & 2020-04-27T20:17:09.139 & 6.45 & 0.71 \\
            & 2020-04-27T20:19:23.068 & 10.33 & 0.31 \\
            & 2020-04-27T20:25:53.415 & 3.88 & 1.55 \\
            & 2020-04-27T21:15:36.398 & 16.89 & 6.47 \\
            & 2020-04-27T21:20:55.561 & 10.79 & 0.96 \\
            & 2020-04-27T21:24:05.936 & 8.40 & 0.42 \\
            & 2020-04-27T21:25:01.037 & 6.50 & 0.39 \\
            & 2020-04-27T21:27:25.367 & 1.91 & 0.47 \\
            & 2020-04-27T21:43:06.346 & 10.18 & 1.66 \\
            & 2020-04-27T21:48:44.062 & 23.29 & 6.59 \\
            & 2020-04-27T21:57:03.989 & 7.93 & 0.23 \\
            & 2020-04-27T21:59:22.528 & 62.26 & 14.88 \\
            & 2020-04-27T22:55:19.911 & 7.41 & 1.97 \\
            & 2020-04-27T23:02:53.488 & 27.20 & 7.10 \\
            & 2020-04-27T23:06:06.135 & 12.17 & 2.02 \\
            & 2020-04-27T23:25:04.349 & 3.49 & 1.75 \\
            & 2020-04-27T23:27:46.293 & 47.50 & 3.23 \\
            & 2020-04-27T23:42:41.143 & 36.42 & 1.93 \\
            & 2020-04-27T23:44:31.818 & 45.00 & 14.49 \\
            & 2020-04-28T00:23:04.763 & 6.11 & 0.69 \\
            & 2020-04-28T00:24:30.311 & 146.14 & 34.49 \\
            & 2020-04-28T00:25:43.946 & 5.95 & 0.25 \\
            & 2020-04-28T00:37:36.160 & 5.48 & 0.63 \\
            & 2020-04-28T00:39:39.565 & 3.47 & 2.29 \\
            & 2020-04-28T00:40:33.077 & 6.21 & 4.28 \\
            & 2020-04-28T00:41:32.148 & 31.49 & 13.76 \\
            & 2020-04-28T00:43:24.784 & 7.71 & 6.52 \\
            & 2020-04-28T00:44:08.210 & 55.55 & 70.82 \\
            & 2020-04-28T00:45:31.098 & 8.04 & 0.86 \\
            & 2020-04-28T00:46:20.179 & 55.15 & 46.93 \\
            & 2020-04-28T00:51:35.912 & 11.16 & 0.77 \\
            & 2020-04-28T00:54:57.448 & 25.87 & 4.45 \\
            & 2020-04-28T00:56:49.646 & 4.39 & 1.44 \\
            & 2020-04-28T01:04:03.146 & 12.42 & 0.77 \\
            & 2020-04-28T02:27:24.905 & 7.69 & 0.20 \\
            & 2020-04-28T03:47:52.140 & 13.99 & 2.00 \\
            & 2020-04-28T04:09:47.317 & 17.18 & 1.89 \\
            & 2020-05-05T03:02:56.033 & 2.58 & 0.42 \\
            & 2020-05-19T18:57:36.305 & 5.15 & 0.17 \\
4th episode & 2021-09-10T00:46:21.000 & 17.50 & 0.63 \\
            & 2021-09-10T01:00:44.000 & 34.71 & 2.36 \\
            & 2021-09-10T01:13:17.000 & 17.34 & 1.11 \\
            & 2021-09-10T01:14:37.000 & 27.50 & 3.63 \\
            & 2021-09-10T01:17:19.000 & 45.96 & 4.78 \\
            & 2021-09-10T01:18:54.000 & 19.35 & 1.78 \\
            & 2021-09-10T01:20:48.000 & 14.41 & 0.98 \\
            & 2021-09-10T01:27:05.000 & 6.25 & 0.90 \\
            & 2021-09-10T02:36:38.000 & 30.00 & 1.32 \\
            & 2021-09-10T02:44:34.000 & 35.00 & 2.80 \\
            & 2021-09-10T02:55:10.000 & 2.79 & 1.07 \\
            & 2021-09-11T03:02:28.000 & 7.78 & 0.56 \\
            & 2021-09-11T13:28:55.000 & 119.64 & 20.10 \\
            & 2021-09-11T15:06:43.000 & 199.73 & 81.49 \\
            & 2021-09-11T15:15:25.000 & 16.59 & 1.46 \\
            & 2021-09-11T15:17:45.000 & 141.90 & 49.38 \\
            & 2021-09-11T15:26:12.000 & 116.36 & 10.24 \\
            & 2021-09-11T15:32:33.000 & 36.21 & 4.49 \\
            & 2021-09-11T15:38:26.000 & 18.06 & 2.60 \\
            & 2021-09-11T16:50:04.000 & 99.77 & 4.39 \\
            & 2021-09-11T17:01:09.675 & 1.18 & 1.80 \\
            & 2021-09-11T17:04:29.740 & 14.59 & 4.30 \\
            & 2021-09-11T17:10:48.619 & 361.19 & 162.51 \\
            & 2021-09-11T20:13:40.000 & 27.33 & 4.70 \\
            & 2021-09-11T20:22:59.000 & 39.75 & 7.95 \\
            & 2021-09-12T00:45:49.000 & 36.25 & 1.45 \\
            & 2021-09-12T06:51:14.000 & 300.37 & 88.91 \\
            & 2021-09-12T07:28:07.000 & 40.80 & 3.59 \\
\enddata
\tablenotetext{a}{the bursts outside of the 4 active episodes that are defined in this study (see Table \ref{tab:wt_log_gaus}). }
% \tablecomments{The Flux in units of $\times 10^{7}$ erg $\cdot$ cm$^{-2}$ s$^{-1}$ and Fluence in units of $\times 10^{7}$ erg $\cdot$ cm$^{-2}$. The time range of the 1st, 2nd, 3rd and 4th episodes is May 2016, June 2016, Apr.-May 2020 and Sept. 2021 respectively (see Table \ref{tab:wt_log_gaus}). }
\end{deluxetable}

\startlongtable
\begin{deluxetable}{lccc}
\tabletypesize{\scriptsize}
\tablecaption{The list of the inferred independent bursts classified by fluence}
\label{tab:independ_burst_fluence}
\tablehead{\colhead{Episode} & \colhead{Time (UTC)} & \colhead{Flux} & \colhead{Fluence}}
\startdata
outside-episodes\tablenotemark{a} & 2014-07-05T09:37:34.484 & 14.55 & 0.32 \\
            & 2015-02-22T19:44:16.881 & 44.00 & 1.76 \\
            & 2015-02-22T19:58:16.658 & 5.96 & 0.62 \\
            & 2015-02-23T05:24:53.876 & 6.78 & 2.17 \\
1st episode & 2016-05-18T09:09:23.800 & 116.29 & 28.84 \\
            & 2016-05-18T10:28:02.817 & 22.79 & 2.37 \\
            & 2016-05-18T15:33:47.010 & 10.71 & 0.60 \\
            & 2016-05-18T17:05:12.813 & 2.73 & 0.24 \\
            & 2016-05-19T11:59:32.632 & 9.11 & 0.51 \\
            & 2016-05-19T12:07:46.594 & 32.03 & 2.05 \\
            & 2016-05-21T21:13:16.242 & 20.25 & 0.81 \\
2nd episode & 2016-06-23T16:49:57.493 & 26.49 & 4.45 \\
            & 2016-06-23T17:47:06.896 & 7.50 & 0.24 \\
            & 2016-06-23T17:55:48.768 & 4.62 & 0.37 \\
            & 2016-06-23T17:58:55.696 & 4.31 & 0.31 \\
            & 2016-06-23T18:22:43.425 & 5.38 & 0.56 \\
            & 2016-06-23T19:36:27.341 & 4.38 & 0.77 \\
            & 2016-06-23T19:37:59.997 & 6.84 & 0.93 \\
            & 2016-06-23T19:51:16.261 & 3.67 & 0.44 \\
            & 2016-06-23T20:06:37.158 & 28.02 & 2.69 \\
            & 2016-06-23T21:15:31.444 & 6.13 & 0.98 \\
            & 2016-06-23T21:16:24.612 & 19.04 & 7.77 \\
            & 2016-06-23T21:20:46.404 & 101.19 & 49.38 \\
            & 2016-06-23T21:23:36.780 & 54.86 & 7.90 \\
            & 2016-06-23T21:30:47.032 & 4.52 & 0.94 \\
            & 2016-06-23T21:38:13.432 & 5.36 & 0.30 \\
            & 2016-06-26T17:50:03.251 & 12.19 & 0.39 \\
outside-episodes\tablenotemark{a} & 2019-11-04T09:17:53.492 & 18.35 & 5.89 \\
            & 2019-11-04T20:13:42.537 & 4.36 & 0.61 \\
            & 2019-11-04T20:29:39.804 & 10.70 & 1.37 \\
            & 2019-11-04T23:48:01.336 & 13.64 & 3.07 \\
3rd episode & 2020-04-27T18:31:05.770 & 3.11 & 0.76 \\
            & 2020-04-27T18:33:53.116 & 7.18 & 0.51 \\
            & 2020-04-27T18:34:05.700 & 47.65 & 20.11 \\
            & 2020-04-27T18:34:46.047 & 16.95 & 3.83 \\
            & 2020-04-27T18:35:05.320 & 49.71 & 5.12 \\
            & 2020-04-27T18:35:46.623 & 21.64 & 1.32 \\
            & 2020-04-27T18:38:20.206 & 8.86 & 0.93 \\
            & 2020-04-27T18:38:53.689 & 7.24 & 1.81 \\
            & 2020-04-27T18:40:15.043 & 2.52 & 1.15 \\
            & 2020-04-27T18:40:32.031 & 4.55 & 6.16 \\
            & 2020-04-27T18:42:40.816 & 11.29 & 0.35 \\
            & 2020-04-27T18:42:50.652 & 25.85 & 8.17 \\
            & 2020-04-27T18:46:08.767 & 18.74 & 3.86 \\
            & 2020-04-27T18:46:39.414 & 5.88 & 9.70 \\
            & 2020-04-27T18:47:05.754 & 76.00 & 11.78 \\
            & 2020-04-27T18:48:38.675 & 6.95 & 1.69 \\
            & 2020-04-27T18:49:28.034 & 18.56 & 6.83 \\
            & 2020-04-27T18:55:44.155 & 36.81 & 2.65 \\
            & 2020-04-27T18:57:35.574 & 8.12 & 0.78 \\
            & 2020-04-27T18:58:45.533 & 4.04 & 0.78 \\
            & 2020-04-27T19:37:39.328 & 5.37 & 3.89 \\
            & 2020-04-27T19:43:44.537 & 88.12 & 38.42 \\
            & 2020-04-27T19:45:00.478 & 5.05 & 0.51 \\
            & 2020-04-27T19:55:32.325 & 5.61 & 0.23 \\
            & 2020-04-27T20:01:45.681 & 10.52 & 5.08 \\
            & 2020-04-27T20:07:20.319 & 2.07 & 0.79 \\
            & 2020-04-27T20:13:38.263 & 9.45 & 0.52 \\
            & 2020-04-27T20:14:51.396 & 27.06 & 1.38 \\
            & 2020-04-27T20:15:20.583 & 148.03 & 189.77 \\
            & 2020-04-27T20:17:09.139 & 6.45 & 0.71 \\
            & 2020-04-27T20:17:50.343 & 1.55 & 2.20 \\
            & 2020-04-27T20:20:44.640 & 2.21 & 0.74 \\
            & 2020-04-27T20:21:51.841 & 2.82 & 1.63 \\
            & 2020-04-27T20:25:53.415 & 3.88 & 1.55 \\
            & 2020-04-27T21:15:36.398 & 16.89 & 6.47 \\
            & 2020-04-27T21:20:55.561 & 10.79 & 0.96 \\
            & 2020-04-27T21:24:05.936 & 8.40 & 0.42 \\
            & 2020-04-27T21:27:25.367 & 1.91 & 0.47 \\
            & 2020-04-27T21:43:06.346 & 10.18 & 1.66 \\
            & 2020-04-27T21:48:44.062 & 23.29 & 6.59 \\
            & 2020-04-27T21:57:03.989 & 7.93 & 0.23 \\
            & 2020-04-27T21:59:22.528 & 62.26 & 14.88 \\
            & 2020-04-27T22:55:19.911 & 7.41 & 1.97 \\
            & 2020-04-27T23:02:53.488 & 27.20 & 7.10 \\
            & 2020-04-27T23:06:06.135 & 12.17 & 2.02 \\
            & 2020-04-27T23:25:04.349 & 3.49 & 1.75 \\
            & 2020-04-27T23:27:46.293 & 47.50 & 3.23 \\
            & 2020-04-27T23:42:41.143 & 36.42 & 1.93 \\
            & 2020-04-27T23:44:31.818 & 45.00 & 14.49 \\
            & 2020-04-28T00:23:04.763 & 6.11 & 0.69 \\
            & 2020-04-28T00:24:30.311 & 146.14 & 34.49 \\
            & 2020-04-28T00:37:36.160 & 5.48 & 0.63 \\
            & 2020-04-28T00:39:39.565 & 3.47 & 2.29 \\
            & 2020-04-28T00:40:33.077 & 6.21 & 4.28 \\
            & 2020-04-28T00:41:32.148 & 31.49 & 13.76 \\
            & 2020-04-28T00:43:24.784 & 7.71 & 6.52 \\
            & 2020-04-28T00:44:08.210 & 55.55 & 70.82 \\
            & 2020-04-28T00:45:31.098 & 8.04 & 0.86 \\
            & 2020-04-28T00:46:00.034 & 7.58 & 6.05 \\
            & 2020-04-28T00:46:20.179 & 55.15 & 46.93 \\
            & 2020-04-28T00:46:43.072 & 7.59 & 3.82 \\
            & 2020-04-28T00:47:57.536 & 5.77 & 1.50 \\
            & 2020-04-28T00:48:44.833 & 4.24 & 1.88 \\
            & 2020-04-28T00:51:35.912 & 11.16 & 0.77 \\
            & 2020-04-28T00:54:57.448 & 25.87 & 4.45 \\
            & 2020-04-28T00:56:49.646 & 4.39 & 1.44 \\
            & 2020-04-28T01:04:03.146 & 12.42 & 0.77 \\
            & 2020-04-28T02:27:24.905 & 7.69 & 0.20 \\
            & 2020-04-28T03:47:52.140 & 13.99 & 2.00 \\
            & 2020-04-28T04:09:47.317 & 17.18 & 1.89 \\
            & 2020-05-05T03:02:56.033 & 2.58 & 0.42 \\
            & 2020-05-19T18:57:36.305 & 5.15 & 0.17 \\
4th episode & 2021-09-10T01:00:44.000 & 34.71 & 2.36 \\
            & 2021-09-10T01:13:17.000 & 17.34 & 1.11 \\
            & 2021-09-10T01:14:37.000 & 27.50 & 3.63 \\
            & 2021-09-10T01:17:19.000 & 45.96 & 4.78 \\
            & 2021-09-10T01:18:54.000 & 19.35 & 1.78 \\
            & 2021-09-10T01:20:48.000 & 14.41 & 0.98 \\
            & 2021-09-10T01:27:05.000 & 6.25 & 0.90 \\
            & 2021-09-10T02:36:38.000 & 30.00 & 1.32 \\
            & 2021-09-10T02:44:34.000 & 35.00 & 2.80 \\
            & 2021-09-10T02:55:10.000 & 2.79 & 1.07 \\
            & 2021-09-11T03:02:28.000 & 7.78 & 0.56 \\
            & 2021-09-11T13:28:55.000 & 119.64 & 20.10 \\
            & 2021-09-11T15:06:43.000 & 199.73 & 81.49 \\
            & 2021-09-11T15:15:25.000 & 16.59 & 1.46 \\
            & 2021-09-11T15:17:45.000 & 141.90 & 49.38 \\
            & 2021-09-11T15:26:12.000 & 116.36 & 10.24 \\
            & 2021-09-11T15:32:33.000 & 36.21 & 4.49 \\
            & 2021-09-11T15:38:26.000 & 18.06 & 2.60 \\
            & 2021-09-11T16:50:04.000 & 99.77 & 4.39 \\
            & 2021-09-11T17:01:09.675 & 1.18 & 1.80 \\
            & 2021-09-11T17:04:29.740 & 14.59 & 4.30 \\
            & 2021-09-11T17:10:48.619 & 361.19 & 162.51 \\
            & 2021-09-11T20:13:40.000 & 27.33 & 4.70 \\
            & 2021-09-11T20:22:59.000 & 39.75 & 7.95 \\
            & 2021-09-12T00:45:49.000 & 36.25 & 1.45 \\
            & 2021-09-12T07:28:07.000 & 40.80 & 3.59 \\
\enddata
\tablenotetext{a}{the bursts outside of the 4 active episodes that are defined in this study (see Table \ref{tab:wt_log_gaus}). }
% \tablecomments{The Flux in units of $\times 10^{7}$ erg $\cdot$ cm$^{-2}$ s$^{-1}$ and Fluence in units of $\times 10^{7}$ erg $\cdot$ cm$^{-2}$. The time range of the 1st, 2nd, 3rd and 4th episodes is May 2016, June 2016, Apr.-May 2020 and Sept. 2021 respectively (see Table \ref{tab:wt_log_gaus}). }
\end{deluxetable}

\clearpage
\bibliography{main}{}
\bibliographystyle{aasjournal}

\end{document}